\documentclass[prd,twocolumn,nofootinbib]{revtex4}
 \usepackage{graphicx}
\usepackage{graphics,epsfig}
 \usepackage{amssymb}
 \usepackage{tabularx}
 \usepackage{ulem}
 \usepackage{multirow}
  \usepackage{color}
  \usepackage{subfigure}
  \usepackage{subfig}
 \usepackage{amsmath}
 \usepackage{multirow}
 \def\be{\begin{equation}}
\def\ee{\end{equation}}
 \def\bi{\begin{itemize}}
 \def\ei{\end{itemize}}
  \def\ben{\begin{enumerate}}
\def\een{\end{enumerate}}
  \def\bt{\begin{tabular}}
\def\et{\end{tabular}}
\def\bc{\begin{center}}
\def\ec{\end{center}}

\def\gam{\gamma}
\def\bea{\begin{eqnarray}}
\def\eea{\end{eqnarray}}
\newcommand\ST{\rule[-1em]{0pt}{2.5em}}

\def\fr{\frac}

\begin{document}

\input{epsf}

\title{Cosmological Implications of the Effective Field Theory of Cosmic Acceleration
}
\author { Eva-Maria Mueller $^{1}$, Rachel Bean $^{2}$,   Scott Watson $^{3}$}
\affiliation{$^1$ Department of Physics, Cornell University, Ithaca, NY 14853, USA, 
\\ $^2$ Department of Astronomy, Cornell University, Ithaca, NY 14853, USA, 
\\$^{3}$  Department of Physics, Syracuse University, Syracuse, NY 13244.}
\label{firstpage}

\begin{abstract}
We consider cosmological constraints arising from the background expansion history on the effective field theory of cosmic acceleration, a theoretical framework that allows for a unified way to classify both models of dark energy  and modified gravity within the linear regime. In the Einstein frame, the most general action for the background can be written in terms of a canonical scalar field which is non-minimally coupled to matter.  The leading corrections to the action are expressible through a quartic kinetic term, and scalar couplings to a Gauss-Bonnet curvature term and the Einstein tensor.  We determine the implications of the terms in this general action for the predicted  expansion history  in the context of dynamical attractors. We find that each modifies the matter dominated and/or accelerative eras in ways that allow us to place cosmological constraints on them. We present current constraints on the effective action using the latest Type Ia supernovae, Cosmic Microwave Background, and Baryonic Acoustic Oscillation data. This includes finding that  the scalar field EFT with a coupled Gauss-Bonnet term and the data are significantly discrepant. 
\end{abstract}

\maketitle

%%%%%%%%%%%%%%%%%%%%%%%%%%%%%%%%%%%%%%%%%%%%%%
\section{Introduction}
\label{sec-intro}
%%%%%%%%%%%%%%%%%%%%%%%%%%%%%%%%%%%%%%%%%%%%%%

Current measurements of the expansion history have provided complementary evidence for the universe's recent transition to accelerative expansion. The primary constraints have come from geometric measurements of cosmic distances from Type Ia supernovae luminosity distances \cite{Kowalski:2008ez, Riess:2006fw,Astier:2005qq, 1999ApJ...517..565P,Riess:1998cb,Hicken:2009df,Amanullah:2010vv}, the angular scale of the sound horizon at last scattering imprinted in the Cosmic Microwave Background (CMB) temperature correlations \cite{Komatsu:2010fb, 2003ApJS..148..175S, Spergel:2006hy, Reichardt:2008ay,Nolta:2008ih, Reichardt:2011yv}, and in the distribution of large scale structure through Baryon Acoustic Oscillations (BAO) \cite{Percival:2007yw,Blake:2011en, Anderson:2012sa}

Understanding the underlying cause of this acceleration, under the broad banner of ``dark energy", represents one of the primary challenges facing contemporary astronomy and astrophysics. Though current constraints are entirely consistent with a cosmological constant, $\Lambda$, a dynamically evolving dark energy is by no means ruled out by the data. Theoretically, the coincidence and fine-tuning problems associated with $\Lambda$ have led theorists to consider broader alternative causes for the acceleration. These include introducing a new type of matter, typically in the form of a scalar field, or modifications to General Relativity on large scales. Many models have been suggested in the literature, e.g. quintessence, non-minimal couplings between quintessence and matter, k-essence, $f(R)$-gravity, Brans-Dicke theories, Galileon gravity, Dvali-Gabadadze-Porrati (DGP) gravity, see, for example, \cite{Copeland:2006wr}  for a review of a wide range of dark energy models, and \cite{Tsujikawa:2010zza,Hinterbichler:2011tt} for recent discussions on modified gravity models. 

There has been significant progress in characterizing possible theoretical mechanisms to generate cosmic acceleration. Arguably, however, the improvements in the precision and breadth of observational constraints, of the expansion history and the growth of large scale structure,  are providing remarkable opportunities to learn about the phenomenological properties of dark energy. As the accuracy of cosmological measurements continues to improve at a rapid pace, it is crucial to establish a robust and systematic way to connect data with fundamental theory. In particle physics and condensed matter systems the framework of Effective Field Theory (EFT) has proven to be very successful at this endeavor (for reviews see \cite{Kaplan:2005es,Burgess:2007pt}). The procedure is to construct the most general theory for observables that is compatible with the expected symmetries of the theory.  The terms in the action can then be constrained by a combination of data from experiments, as well as ensuring theoretical self consistency.

The application of the EFT approach to cosmology has recently been carried out for single and multi-field inflation~\cite{Cheung:2007st,Senatore:2010wk}.  In these papers the authors construct the most general action for the cosmological fluctuations around an assumed inflationary background. For inflation, this is reasonable, since apart from establishing a successful period of inflation the key observables are the correlation functions of scalar and metric perturbations. However, the same is not true for present day acceleration where the background expansion itself gives rise to observables, motivating the construction of the EFT of the background. Such an approach was first developed\footnote{For a review of earlier work utilizing EFT methods to address the importance of high energy signatures on inflation we refer the reader to \cite{Avgoustidis:2012yc} and references within.} for inflation by Weinberg \cite{Weinberg:2008hq} and for late-time cosmic acceleration in \cite{Park:2010cw} -- with later work appearing in \cite{Bloomfield:2011wa, Jimenez:2011nn,Jimenez:2012jg,Khosravi:2012qg}.

The goal of these approaches is to systematically categorize all proposals for cosmic acceleration, other than a cosmological constant, so they can be efficiently scrutinized by theory and experiment simultaneously.  To facilitate such an approach a Lagrangian was constructed that could not only account for dark energy models, such as quintessence, but also theories of modified gravity in the linear regime. 
The relevant Lagrangian is that of a scalar-tensor theory non-minimally coupled to gravity, where in the modified gravity case the scalar can be interpreted as the extra degree of freedom arising from the longitudinal component of the graviton.

In \cite{Creminelli:2008wc}, the authors restricted their attention to quintessence and utilized the EFT of the {\it perturbations} around the accelerating background, as was done for inflation in \cite{Cheung:2007st}.  This has the advantage that perturbations about non-linear backgrounds can be considered, but the disadvantage that the acceleration of the background must be assumed {\it a priori}.  Regardless, this type of approach is important when considering models beyond the linear regime, and in particular for models of modified gravity where recovering the predictions of General Relativity and consistency with solar system tests requires the EFT of the background to ultimately fail -- e.g. in regions of high density leading to a screening mechanism e.g. \cite{Khoury:2003aq,Jain:2010ka,Hinterbichler:2010es}.  

In addition to the EFT background approach, one may also consider imposing symmetries on non-linear backgrounds to retain a well defined Cauchy problem.  In the EFT approach, this is never a concern since for energies below the cutoff of the theory, the number of degrees of freedom remains fixed, and higher time derivatives never appear in the equations of motion \cite{Weinberg:2008hq}.  However, as shown long ago by Horndeski \cite{1974IJTP...10..363H}, by restricting the operators to be considered in the action it is possible to construct non-linear backgrounds that give rise to only two derivatives acting on fields at the level of the equations of motion.  These models are less general than the EFT approach, but in addition to being able to capture non-linearities they also exhibit interesting self-tuning properties.  The authors of \cite{Charmousis:2011bf} recently studied these actions identifying four terms that are important for classifying these scalar-tensor theories.  Given our EFT approach, these terms are captured at leading order by the action to be considered below, but the full non-linear effects will not be captured.  

A powerful approach to understanding the implications of a given theory is to find dynamical attractor solutions for cosmological evolution. These provide predictions of evolutionary trajectories that a theory will naturally tend towards, largely insensitive to assumptions about initial conditions. Dynamical attractor analyses have allowed inferences about the viability of dark energy theories, in the context of minimally coupled quintessence theories \cite{Ferreira:1997hj, Wetterich:1994bg}, non-minimal couplings in the dark sector \cite{Bean:2008ac}, and $f(R)$ theories \cite{Amendola:2006kh}, and see \cite{Copeland:1997et} for a rather comprehensive review of dynamical attractors in dark energy. There are ways to evade dynamical attractors, for example by choosing forms of self-interactions that, by construction, do not allow attractors \cite{Hu:2007nk}. While such selections provide a proof of principle, they explicitly require a fine-tuning by the specificity of their form, which the generality of dynamical attractors works to avoid.

In this paper we consider the implications  for dynamical attractors in the background expansion history of an effective theory of cosmic acceleration expressed in terms of the low-energy effective action proposed by Park, Watson, and Zurek  \cite{Park:2010cw} and Bloomfield and Flanagan \cite{Bloomfield:2011wa}. Our results here apply strictly to the linear regime. There has been recent work looking at  observational implication of the non-linear EFT of perturbations \cite{Battye:2012eu, Pearson:2012kb,Baker:2012}.  

 In section \ref{sec-eft} we outline the effective field theory action which we consider in the paper and discuss the origins of the contributing terms and the resulting equations of motion.  In section \ref{sec-attractor} we present dynamical attractor solutions for the components of the effective theory and consider the analytical and numerical implications for the background expansion history. In section \ref{sec-constraints} we outline the cosmological data sets we use to constrain the effective theory and summarize our results. We bring together our findings and discuss implications for future work in section \ref{sec-conclusions}.

%%%%%%%%%%%%%%%%%%%%%%%%%%%%%%%%%%%
\section{Effective Field Theory for Dark Energy}
\label{sec-eft}
%%%%%%%%%%%%%%%%%%%%%%%%%%%%%%%%%%%%%

Working in the Einstein frame, we consider the effective theory for the background with leading scalar and gravitational corrections \cite{Bloomfield:2011wa, Park:2010cw}
\bea
S &=& \int d^4 x \sqrt{-g}  \left\{ \frac{M_p^2}{2} R - \frac{1}{2} (\nabla \phi)^2 - V(\phi) \right\} 
    \nonumber \\
&& + \int d^4 x \sqrt{-g}   \left\{ f_{quartic} (\phi)(\nabla \phi)^4 + f_{curv}(\phi)G^{\mu \nu} \nabla_\mu \phi \nabla_\nu \phi\right.
\nonumber \\
&& \left.+f_{GB} (\phi) \left( R^2 - 4 R^{\mu \nu} R_{\mu \nu} + R_{\mu \nu \sigma \rho} R^{\mu \nu \sigma \rho} \right) \right\} 
\nonumber \\
&&+ S_{\rm m} \left[ \Omega(\phi,\nabla\phi) g_{\mu \nu}, \psi_{\rm m} \right] 
\label{action}
\eea
where $M_p=1/\sqrt{8\pi G}=2.43 \times 10^{18}$ GeV is the reduced Planck mass and $\psi_m$ are the standard model and dark matter fields.  The first line gives the leading order terms, a canonical scalar field with an arbitrary potential $V(\phi)$. The second line gives the leading derivative corrections: a quartic kinetic term and a direct coupling between the scalar gradient and the Einstein tensor. The third line is a Gauss-Bonnet (GB) curvature term.  The last line describes the non-minimally coupling between the scalar field and matter in the Einstein frame, 
with
\bea
\Omega (\phi,\nabla\phi) &=& e^{\alpha(\phi)} \left( 1 + f_{kin}(\phi)
    (\nabla \phi)^2 \right)
\eea
where $\alpha$ is the leading order coupling and a term proportional to $f_{kin}$ provides the next order correction.  The coefficients $f_{quart}$, $f_{curv}$,  $f_{GB}$ and $f_{kin}$, along with $V$ and $\alpha$, are all arbitrary functions of the scalar field.  We note that this implies that the GB term is no longer a total derivative. 

 We focus here on terms in the effective theory that may  influence cosmic evolution in the matter dominated and late time accelerative eras.   We do not consider higher order corrections to the energy-momentum tensor, ${T^{\mu}}_{\nu}=\mbox{diag}(-\rho,P,P,P)$ where $\rho$ is the cosmic matter density and $P$ is the isotropic pressure that can arise in the a general EFT. Specifically we neglect operators such as those proportional to $({T^\mu}_\mu)^2=(-\rho+3P)^2$ or $T^{\mu\nu}T_{\mu\nu}=(\rho^2+3P^2)$, since these types of corrections would most likely lead to effects at all times, and be particularly relevant in the high density regime during radiation domination.

The power of the EFT approach is that data from observations can be used to constrain the possible values of the parameters, while at the same time theoretical consistency places additional constraints to eliminate regions of the parameter space. As an example of the latter, it is well known that radiative stability of the scalar field for quintessence puts strong constraints on the form of the scalar potential \cite{Kolda:1998wq}.
While the coefficients in the action are arbitrary functions of the scalar field, it is useful to expand these functions in powers of the cutoff of the effective theory.  The cutoff is a dimensionful scale that characterizes the higher energy, microscopic physics that has been integrated out as our observational interest passes to lower energy scales.  For example, for the application to dark energy this scale could result from integrating out electrons. 
We note, that given the cosmological background is dynamic (characterized by the Hubble parameter, $H(t)$), this can lead to several scales besides the masses of heavy particles going into determining the effective cutoff(s) of the theory -- see e.g. \cite{Burgess:2009ea}. Lacking a knowledge of the high energy completion here, in this paper we will study the implications for each leading correction by expanding functions with time derivatives of the scalar scaled by powers of the Hubble scale, whereas the scalar field will be suppressed by powers of the Planck mass.  

We emphasize that since we are considering the EFT of the background (not the perturbations) our expansion must always respect the hierarchical structure of the terms, i.e. higher order corrections must always be subleading to lower order terms.  This is particularly important to keep in mind in regards to the GB term, as this type of term is often considered with a large pre-factor, which can lead to a number of instabilities\footnote{For a discussion of the pathologies associated with large GB terms and the difference with the EFT approach we refer the reader to \cite{Weinberg:2008hq}.}.  Here we will follow the EFT approach and treat the GB term as the first in a serious of higher derivative corrections.  Indeed, the procedure followed in \cite{Park:2010cw,Bloomfield:2011wa} for arriving at the general action (\ref{action}) requires that this term remain small for the reduction procedure that was implemented there to be valid. We also emphasize that the presence of the scalar dependent coefficient $f_{GB}$ implies that the GB term is {\it not} purely topological and so can play an important role in the dynamics.

Our cutoffs, chosen to scale with powers of the Hubble and Planck scales, represent the most optimistic case for observations, and therefore the strongest constraints for theories.  For larger cutoffs the observational effects of higher dimensional operators for dark energy and modified gravity are known to be negligible.  Models of ghost condensation offer an example, where there the cutoff is typically taken to be of order a few GeV, and so the evolution is indistinguishable from a cosmological constant \cite{ArkaniHamed:2003uy}.  For the EFT of quintessence, a GeV scale cutoff implies a similar story, and all models remain observationally indistinguishable from ordinary, vanilla quintessence. Thus, here we focus on the lowest possible cutoff consistent with the effective theory, leaving the important question of the UV completion to future work.
 
Given these considerations we take the couplings to terms involving scalar gradients  to be of the form:
\bea
f_{quart} &=& \frac{F_q}{M_p^2H^2} \label{Fqdef}
\\
f_{curv} &=& \frac{F_c}{H^2} \label{Fcdef}
\\
f_{kin} &=&\frac{F_{k}}{M_p^2 H^2} \label{Fkdef}
\eea
where $F_q, F_c$, $F_k$ are constants. Following common examples from the literature, we assume exponential couplings and potentials for the interactions not involving scalar gradients,
\bea
V &=& V_0 \exp\left(-\lambda\frac{\phi}{M_p}\right)
\\
e^{\alpha} &=& \exp\left(-2 Q\frac {\phi}{M_p}\right)
\\
f_{GB} &=& F_0\exp\left(\mu\frac{\phi}{M_p}\right)
\eea
with $V_0$, $\lambda$, $Q$,  $F_0$ and $\mu$  constant.
Our   definitions of $\lambda$, $\mu$ and $Q$ are chosen to be consistent with previous work, such as \cite{Ferreira:1997hj,Tsujikawa:2006ph}, and  $Q=\sqrt{2/3}C$ in \cite{Bean:2008ac}. While we consider an exponential potential in this analysis,  power law potentials are also commonly considered, and though the findings differ in the details, typically they yield broadly comparable results in relation to the constraints on the coupling $Q$ e.g. \cite{Bean:2008ac,Pettorino:2012ts}.
 
%%%%%%%%%%%%%%%%%%%%%%%%%%%%%%%%%%%%%%
\subsection{The equations of motion}
%%%%%%%%%%%%%%%%%%%%%%%%%%%%%%%%%%%%%%

We assume a Friedmann Robertson Walker (FRW) metric,
\bea
ds^2 = -dt^2 + a^2dx^2
\eea 
where $t$ is physical time and $a$ is the scale factor and $H=\dot{a}/a$.
Given a homogeneous scalar field, the Einstein field equations and energy-momentum conservation equations for the scalar and matter give rise to consistent general equations of motion given by
the Friedmann equation,
\bea
3M_p^2{H}^{2} &=& {\rho}_{m}(\phi)+{\rho}_{\gam}+\frac{1}{2}\dot\phi^2+{V}(\phi)+ 3f_{quart}\dot\phi^4
\nonumber \\
&& + 9 f_{curv} H^2 \dot\phi^2+ 24 \dot{\phi} f_{GB}'H^3 ,  \label{eq:FE}
\eea 
the acceleration equation,
\bea
-3 && \hspace{-0.25in}M_p^2H^2 \left[\frac{2}{3}\frac{\dot{H}}{H^2}\left(1- f_{curv} \frac{\dot\phi^2}{M_p^2}\right)+1\right] \nonumber \\
 \hspace{-0.3in}&=& \frac{1}{3} {\rho}_{\gamma}+ \frac{1}{2}\dot{\phi}^2 
-V(\phi)+ f_{quart}\dot\phi^4\nonumber
\\
 \hspace{-0.3in} &-&f_{curv} (3 H^2 \dot\phi^2 + 4  H \dot\phi \ddot{\phi}) - 2 f_{curv}^\prime H \dot\phi^3\nonumber
\\
 \hspace{-0.3in} & -& 8 H^3 f_{GB}' \dot{\phi}\left(\frac{f_{GB}''\dot\phi}{f_{GB}'H}+\frac{\ddot{\phi}}{\dot\phi H} +2\frac{\dot{H}}{H^2}+2 \right), \hspace{0.75cm}
\label{eq:accel}
\eea
the scalar field equation,
\bea
&&(1+12f_{quart}\dot\phi^2)\ddot{\phi} +  \left(3+12f_{quart}\dot\phi^2\right) {H} \dot{\phi} + V' = 
\nonumber \\
&&  \rho_m\left(\frac{Q}{M_p} +\frac{f_{kin}^\prime \dot\phi^2 + 2 f_{kin}  \ddot\phi}{2(1-f_{kin} \dot\phi^2)} \right)-3 f_{quart}^\prime \dot\phi^4\nonumber
\\
&& -24 f_{GB}'H^4 \left(\frac{\dot{H}}{ H^2}+1\right)\nonumber
\\
&& - f_{curv} (6H^2 \ddot\phi+12H \dot{H} \dot\phi+18H^3 \dot\phi)-3 f_{curv}^\prime H^2 \dot\phi^2,  \hspace{0.75cm}\label{eq:phi}
\eea
and the matter fluid equation
\bea
\dot\rho &=&-3H(\rho+P) 
\nonumber
\\
&&- \left(\frac{Q}{M_p}\dot\phi + \frac{f_{kin}^\prime \dot\phi^3 + 2 f_{kin} \dot\phi \ddot\phi}{2(1-f_{kin} \dot\phi^2)} \right)(\rho-3P), \hspace{0.75cm}\label{eq:fluid}
\eea
where primes denote derivatives with respect to the scalar field and dots denote derivatives with respect to physical time. $\rho$ and  $P$ are functions of $\phi$ through $Q$ and $F_{kin}$, unless conformally coupled.

While it is convenient to perform the dynamical attractor analysis in the Einstein frame, comparisons with observations are  performed in the Jordan frame, as typically cosmological redshifts in data assume baryonic matter is minimally coupled, and not subject to  fifth forces due to a direct scalar coupling. To transform into the Jordan frame we use the conformal transformation to remove the coupling $\Omega$ from the metric, with $\tilde{g}_{\mu \nu} = \Omega g_{\mu \nu}$,   $d\tilde{t}=\Omega^{-1/2}dt$, $ \tilde{a} =\Omega^{-1/2}a$ and matter variables  transforming as $\tilde{P}= \Omega^{-2} P$, and $\tilde\rho = \Omega^{-2}\rho$, where tildes denote Jordan frame variables.

%%%%%%%%%%%%%%%%%%%%%%%%%%%%%%%%%%%%
\section{Dynamical Attractor Solutions}
\label{sec-attractor}
%%%%%%%%%%%%%%%%%%%%%%%%%%%%%%%%%%%%
In this section we use the background equations of motion to look for the stationary `attractor' solutions during the matter and late time accelerative eras. We will consider the impact of the higher order operators arising in the effective field theory  on the cosmological dynamics compared with the behavior predicted by the leading order action.

%--------------------------------------------
\subsection{Leading order terms}
\label{sec-lead}
%--------------------------------------------

The standard dynamical solutions for a non-minimally coupled, canonical scalar field \cite{Ferreira:1997hj, Wetterich:1994bg,Copeland:1997et} are commonly written using dimensionless variables, see for example \cite{Copeland:1997et},
\bea
x&=& \fr{1}{M_p H}\fr{\dot{\phi}}{\sqrt{6}}, \ \ \ y = \fr{1}{M_p H}\fr{\sqrt{V}}{\sqrt{3}}, \ \ \ z = \fr{1}{M_p H}\fr{\sqrt{\rho_{\gam}}}{\sqrt{3}}. \hspace{0.75cm}
\eea
The Friedmann equation gives the fractional matter density, $\Omega_m$, in terms of these variables 
\bea
\Omega_m(a) =\frac{\rho_{m}}{3M_p^2H^{2}} &=& 1-x^{2}-y^{2}-z^{2}. \label{eq:Om}
\eea
The acceleration equation gives the effective Einstein frame equation of state parameter, $w_{E}$,
\bea
-\frac{\dot{H}}{H^2}
= \frac{3}{2}(1+w_{E})&=&\frac{3}{2}\left(1+x^2-y^2+\frac{1}{3}z^2\right),
\eea
and an effective scalar equation of state parameter,
\bea
w_{\phi}&=& \frac{w_{E}-\frac{1}{3}z^2}{x^2+y^2}=\frac{ x^{2}-y^{2}}{x^2+y^2}.
\eea

The scalar and matter fluid equations give the evolution for $x,y$, and $z$,
\bea
\frac{dx}{d\ln a} &=&-x\left(3+\frac{\dot{H}}{H^2}\right)+\frac{\sqrt{6}}{2}\lambda y^2 + \frac{\sqrt{6}}{2}Q\Omega_m(a), \label{eq:xprime} \hspace{0.75cm}
\\
\frac{dy}{d\ln a}  &=&-y\left(\frac{\sqrt{6}}{2}\lambda x+\frac{\dot{H}}{H^2}\right), \label{eq:yprime}
\\
\frac{dz}{d\ln a} &=&-z\left(2+\frac{\dot{H}}{H^2}\right).  \label{eq:zprime}
\eea

The dynamical attractors are given by the static solutions to these equations $dx/d\ln a=dy/d\ln a=dz/d\ln a=0$. 

We  can write a general expression for the effective Jordan equation of state, $w_J$, during attractor-driven epochs, in terms of the  coupling and $w_E$, using the  conformal transformation,
\bea
3(1+w_{J})&=&\frac{3(1+w_{E})-2\sqrt{6}Qx}{1- \sqrt{6}Qx}  \label{eq:weffJ},
\eea
where $x$ is the fractional scalar kinetic energy component in the attractor regime. This relationship holds true even when the non-relativistic matter density, and its effect on the fluid equations, is negligible, and is well-defined as long as $x\neq 1/\sqrt{6}Q$.

Two primary matter era attractor solutions arise: one dependent on the potential (MAT-$\lambda$), and one wholly determined by the non-minimal coupling (MAT-Q). A single potential-driven late time accelerative  attractor   exists (ACC-$\lambda$).

%====================================================
\begin{table*}[t]
\begin{tabular*}{0.676\textwidth}{|l||c|c|c|c|c|c|} 
\hline
Attractor &  $x$ & $y$ & $\Omega_\phi$ & $w_\phi$ & $w_E$ &$w_J$
\\ \hline\hline
RAD-$\lambda$ 
& $\frac{2\sqrt{6}}{3\lambda}$
& $\frac{2\sqrt{3}}{3 \lambda}$
&$\frac{4}{\lambda^2}$
&1/3
&1/3
&$\frac{1}{3}\frac{(\lambda+4Q)}{(\lambda-4Q)}$
\ST \\ \hline
RAD-null 
&  0
&0
&0
&--
&1/3
&1/3
\ST \\ \hline\hline
MAT-$\lambda$ 
&  $\sqrt{\frac{3}{2}}\frac{1}{\lambda-Q} $
&$ \frac{\sqrt{\frac{3}{2}-Q(\lambda-Q)}}{\lambda-Q}$
&$\frac{3}{(\lambda-Q)^2}-\frac{Q}{\lambda-Q}$
&$\frac{Q(\lambda-Q)}{Q(\lambda-Q)-3}$
&$\frac{Q}{\lambda-Q} $
& $\frac{2Q}{\lambda-4Q}$ 
\ST \\ \hline
MAT-Q&$\sqrt{\frac{2}{3}}Q$ 
& $0  $
& $\frac{2Q^2}{3} $
& $ 1 $
& $\frac{2Q^2}{3} $
& $\frac{4Q^2}{3(1-2Q^2)}$
\ST \\ \hline\hline
ACC-$\lambda$ 
& $\frac{\lambda}{\sqrt{6}} $
& $\sqrt{1- \frac{\lambda^2}{6}}$  
& $1$ 
& $-1+\frac{\lambda^2}{3}  $
& $-1+\frac{\lambda^2}{3}$ 
&$ -1+\frac{\lambda(\lambda-2 Q )}{3(1-Q\lambda)} $
\ST \\ \hline
 \end{tabular*}
  \caption{ Table summarizing the properties of the  principal dynamical attractors arising from the leading order terms in the effective action. They consist of two radiation era attractors, RAD-$\lambda$ and RAD-null, two matter era attractors, MAT-$\lambda$ and MAT-Q, and one accelerative era attractor, ACC-$\lambda$. We give the values of the dimensionless scalar field dynamical variables, $x$ and $y$,  the Einsten frame  fractional energy density, $\Omega_\phi$ and equation of state, $w_{\phi}$, for the scalar field, and the overall effective equation of state in the Einstein, $w_E$, and Jordan frame, $w_J$.}
    \label{tab:attractor}
\end{table*}
%====================================================
For a minimal coupled scalar field, when $Q=0$, the MAT-$\lambda$ attractor is the main cosmological scaling solution \cite{Ferreira:1997hj, Wetterich:1994bg}.  In this case, the scalar field evolves with the same equation of state, $w$, as the dominant matter component, with the attractive property that the radiation and matter dominated eras evolve as in $\Lambda$CDM. To satisfy this scaling solution the scalar must be subdominant, $\Omega_{\phi}<\Omega_{m}$ requiring $\lambda>\sqrt{6}$. Nucleosynthesis puts a stronger lower bound on $\lambda$ in the RAD-$\lambda$ attractor; the non-zero scalar density increases the expansion rate, altering the primordial abundances. For example \cite{Bean:2001wt}  reported a dark energy density of $\Omega_\phi<0.09$ which translates into the constraint $\lambda>6.5$.  One cannot, however, simultaneously realize a cosmological solution that flows from this matter era attractor to an accelerative expansion at late times,  as acceleration with the ACC-$\lambda$ attractor requires $\lambda<\sqrt{2}$.  A simple exponential dark energy model cannot  provide a viable matter and accelerative era without an degree of freedom, such as a double potential or a feature in the potential \cite{Albrecht:1999rm}. 

In the non-minimally coupled scenario, an alternative matter era attractor, MAT-Q, is also present. The benefit of this  attractor is that it can arise from the RAD-null attractor, evading the BBN constraints on $\lambda$, and transition to an accelerative ACC-$\lambda$ attractor at late times (for appropriate choice of $\lambda$). It runs into problems observationally, however, because during the matter era $w_{J}\neq 0$. This was first highlighted in the context of $f(R)$ theories   \cite{Amendola:2006kh}, for which $Q=1/\sqrt{6}$, $w_E=1/9$ and $w_J=1/3$ during the matter dominated era, showing that dynamical attractors for such theories are inconsistent with data. The MAT-$\lambda$ attractor also deviates from the dominant background equation of state when $Q=0$, putting it in tension with data. 

The accelerative expansion, in ACC-$\lambda$ is enhanced as Q is increased for $0<Q<1/\lambda$, or $Q>(3-\lambda^2)/\lambda$ for $\lambda<\sqrt{2}$. In the case $\lambda > \sqrt{3}$ the coupling Q must be larger than $Q>1/ \lambda$. This could expand the range of $\lambda$ that can give rise to acceleration.  

%--------------------------------------------
\subsection{Gauss-Bonnet term}
\label{section_GB}
%--------------------------------------------

If uncoupled to the scalar, the Gauss-Bonnet (GB) term is a topological term which  would play no role in the cosmic dynamics. If coupled, however, it will impact  the expansion history  \cite{Tsujikawa:2006ph, Koivisto:2006ai}. 

We introduce an extra dimensionless variable for the GB term
\bea
v &\equiv & \frac{8f_{GB}^\prime H^2}{M_p} .
\eea
 The modified Friedmann equation (\ref{eq:FE}) yields a matter density 
\bea
\Omega_m(a) &=& 1-x^{2}-y^{2} - z^2 -  \sqrt{6}xv   
\eea
and the acceleration equation gives 
\bea
&&-\frac{2}{3}\frac{\dot{H}}{H^2}\left[1-v\left(\sqrt{6}x-\frac{3}{2} v\right)\right] = 1+\frac{1}{3}z^2+x^2-y^2
\nonumber
\\ && \hspace{1.5cm}-v\left( \lambda y^2 +Q\Omega_m(a)+2\mu x^2-v-\sqrt{\frac{2}{3}} x\right).\hspace{0.75cm}
\eea
The fluid equations for $y$ and $z$ have the same forms as in (\ref{eq:yprime}) and (\ref{eq:zprime}), while the dynamical equation for the kinetic term, $x$, and the GB term are given by
\bea
\frac{d x}{d\ln a}&=&-x\left(3+\frac{\dot{H}}{H^2}\right)
\nonumber
\\ && +\frac{\sqrt{6}}{2}\left[\lambda y^2 + Q\Omega_m(a) -  v \left( 1+\frac{\dot{H}}{H^2} \right) \right],
\\
\frac{dv}{d\ln a} &=& 2 v \left( \frac{\sqrt{6}}{2} \mu x + \frac{\dot{H}}{H^2}  \right).
\eea

The matter dominated era allows static solutions with $v=0$, so that the GB term is negligible and attractors are the same for the leading terms: MAT-$\lambda$ and MAT-$Q$ given in Table \ref{tab:attractor}. The difference in including the GB term occurs in the late time accelerative era. An alternative attractor  exists, which is independent of the coupling $Q$, given by
\bea
\begin{array}{|c|c|c|c|c|c|c|c|} 
\hline 
\mathrm{Attractor} & x & y & v & \Omega_\phi & w_\phi & w_E & w_J
\\ \hline
\mathrm{ACC-GB} & 0 & 1& \lambda & 1 & -1& -1 & -1
\\ \hline
\end{array} \label{tab:GBattractor}
\eea
This allows cosmological evolution following the MAT-$\lambda$, with a scaling attractor in the matter dominated era, to transition to a Gauss-Bonnet induced, de-Sitter solution at late times. Note that the accelerative era arises because the gradient in the GB term becomes comparable, and opposite to that of the potential, however the energy density in the GB term, given by $\sqrt{6}xv$, tends to zero in both the matter and de-Sitter epochs.  

  At late times $v=\lambda$, $y=1$ and $x=0$, and $\phi$ is constant. The time at which the  transition from the matter dominated to accelerative era occurs depends on the relative importance of the potential, $V_0$, and Gauss-Bonnet term, $F_0$, in the scalar field equation. We can obtain an approximate relationship between the two by assuming the accelerative attractor is  reached today, with $H=H_0$, \cite{Koivisto:2006ai},
\bea
v &=& \lambda \approx 8f_{GB}'\frac{H_0^2}{M_p}\left.\right|_{\phi=\phi_0} = \frac{-8F_0 \mu H_0^2 }{M_p^2} \exp \left(- \frac{\mu \phi_0}{ M_p} \right) ,\hspace{0.75cm}
\\
y&=&1 \approx \frac{V_0}{3 M_p^2 H_0^2} \exp \left( -\lambda \frac{\phi_0}{M_p} \right).
\eea
This gives an approximate relation to estimate the value of $F_{0}$ required to give acceleration today, in terms of the potential,
 \bea
 F_{0}^{est}=\frac{\lambda M_p^2 }{8 \mu H_0^2 }\left( \frac{3M_p^2H_0^2}{V_0}\right)^{\mu/\lambda}.\label{F0est}
 \eea
 This relation is only a rough estimate, as the transition occurs prior to today and we have not yet reached the pure accelerative era. It gives a sufficiently good starting-point, however, to guide a nuisance parameter in the MCMC analysis as discussed in \ref{sec-mcmc}.

%--------------------------------------------
\subsection{Quartic term}
%--------------------------------------------

%==================================================
\begin{figure*}[!t]
\bc
{\includegraphics[width=0.45\textwidth]{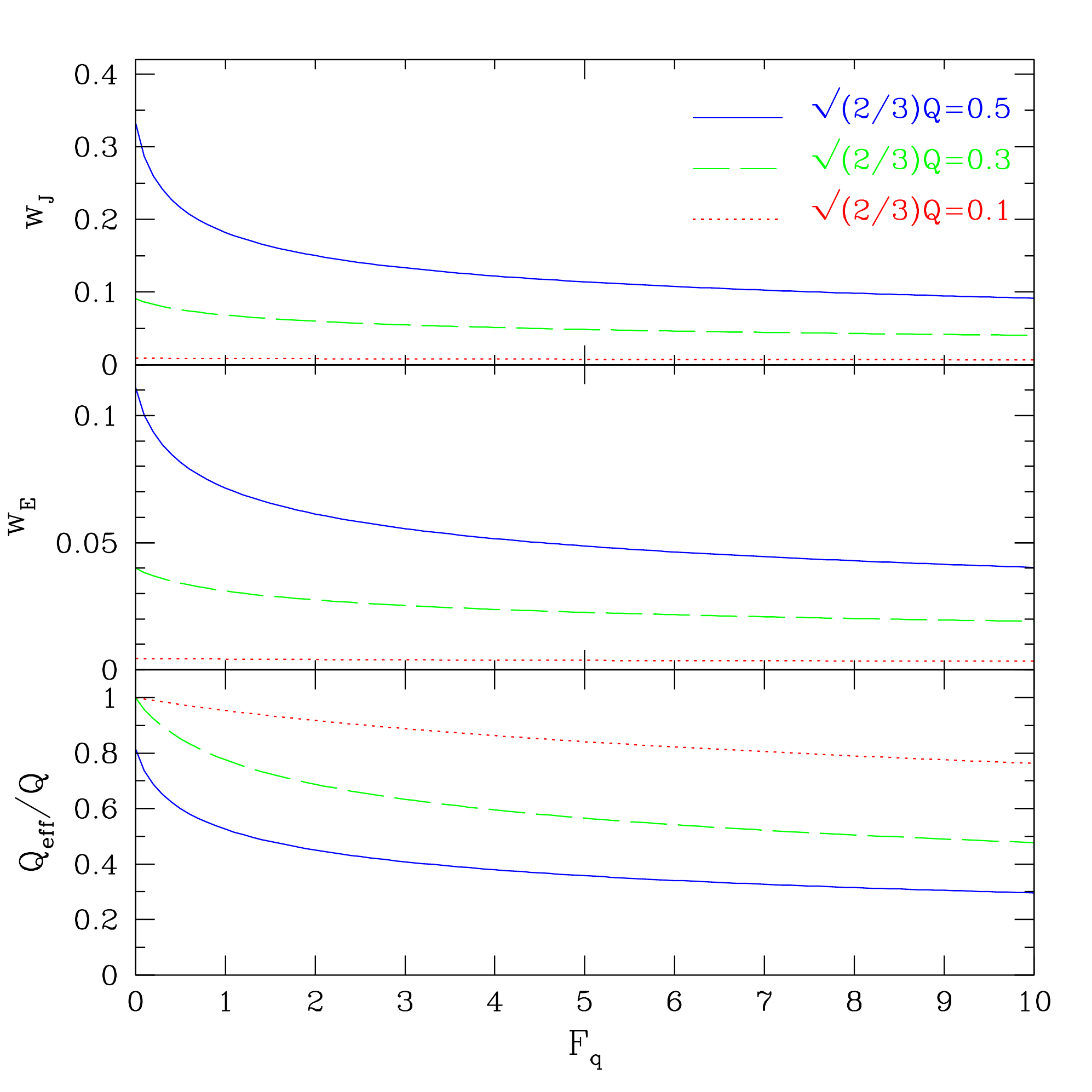}
\includegraphics[width=0.45\textwidth]{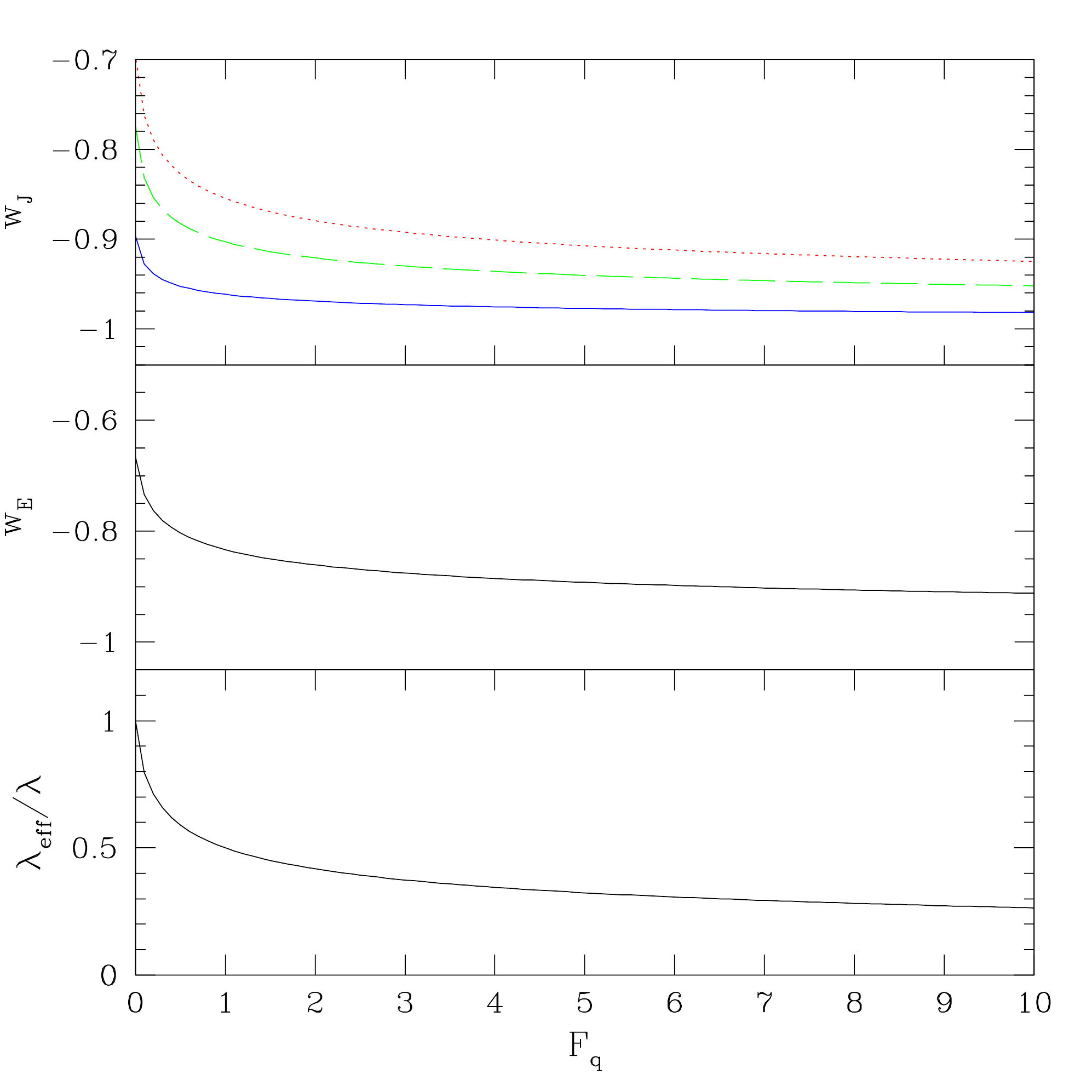}}
\ec
\caption{The implications of the attractor solutions in the matter-dominated [left panel] and accelerative [right panel] eras when  the leading order terms in the action are augmented by the quartic kinetic term parameterized by an amplitude $F_q$. The panels show  [top] the effective equations of state in the Jordan frame,  $w_J$, and [middle]  the Einstein frame, $w_E$ along with [bottom] the effective coupling $Q_{eff}$ [right] and potential exponent $\lambda_{eff}$ [left], relative to their fiducial  values.  Three values of non-minimal couplings $\sqrt{3/2}Q=0.1,0.3,0.5$ are considered. During the accelerative era $Q$ only affects the conformal transformation and Jordan frame equation of state.}
\label{fig:lamQeff}
\end{figure*}
%==================================================

We now consider the implications of an additional quartic term $f_{quart}\dot\phi^4$ in the action, with the coupling $f_{quart}$ parameterized as in (\ref{Fqdef}). The Friedmann equation gives the fractional matter density, $\Omega_m$, in terms of these variables 
\bea
\Omega_m(a) &=& 1-x^{2}-y^{2}-z^{2}-36F_q x^4.
\eea
The acceleration equation becomes
\bea
-\frac{\dot{H}}{H^2}
&=&\frac{3}{2}\left[1+x^2-y^2+\frac{1}{3}z^2+12F_q x^4\right],
\eea
with an effective, overall equation of state
\bea
w_{E} &=&x^2-y^2+\frac{1}{3}z^2+12F_qx^4.
\eea
The fluid equations for $y$ and $z$ have the same forms as in (\ref{eq:yprime}) and (\ref{eq:zprime}), while the dynamical equation for the kinetic term, $x$, is given by
\bea
(1+72F_q x^2)x' &=&-x\left(3+(1+36x^2F_q)\frac{\dot{H}}{H^2}+ 72 F_q x^2\right)
\nonumber
\\ &&+\frac{\sqrt{6}}{2}\lambda y^2 + \frac{\sqrt{6}}{2}Q\Omega_m(a).
\eea

The RAD-null solution during the radiation dominated era is unchanged.  For the $RAD-\lambda$ scaling solution $z$ is changed to $z=\frac{\sqrt{\lambda^4-4 \lambda^2-256 F_q}}{\lambda^2}$ however this just represents an adjustment of the relative contributions of $\Omega_\gamma$ and $\Omega_\phi$ that keeps the effective equation of state  unaltered, $w_E=\frac{1}{3}$.
The same is true for the $MAT-\lambda$ scaling,  the solution for $y$ is altered but  $w_E=\frac{Q}{\lambda-Q}$ is unchanged.

 The coupling in the matter dominated era still admits the MAT-Q attractor, but with a coupling that is dependent on both $Q$ and $F_q$.

 Considering all the terms in $x'=0$ one finds a matter attractor solution, $x_{MAT}$, that differs from the leading case 
\bea
0 &=&1296F_q^2 x_{MAT}^7+144F_q x_{MAT}^5-36\sqrt{6}QF_qx _{MAT}^4
\nonumber
\\ &&+3(1-12F_q)x_{MAT}^3-\sqrt{6}Q x_{MAT}^2 -3x_{MAT}+\sqrt{6}Q.\hspace{0.75cm}
\eea

We can define an effective coupling strength, $Q_{eff}$, based on this attractor solution which would yield the same equation of state in the absence of the quartic term
\bea
Q_{eff}= \sqrt{\frac{3}{2}}x_{MAT}(Q,F_{q}).
\eea
The Einstein frame effective equation of state during the MAT-Q attractor is
 \bea
 w_{E}
 &=&\frac{2}{3}Q_{eff}^2(1+8F_q Q_{eff}^2).
 \eea
While the non-minimal coupling, $Q$, in essence, speeds up the scalar's evolution, and increases $w_E$, the quartic term, $F_q$, has the opposite effect, acting as a resistive force on the scalar and suppressing the kinetic fractional energy density in the MAT era , $x_{MAT}$. The effective equations of state in both the Einstein and Jordan frames are brought closer to $w=0$ as $F_q$ increases. In theory, therefore, this quartic coupling might lessen the tension between non-minimally coupled attractor solutions and constraints on the matter dominated era evolution from CMB distance measurements.

In the accelerative era $y^2=1-x^2-36F_qx^4$, and the attractor solution satisfies
\bea
24F_q x_{ACC}^4+x_{ACC}^2-\frac{1}{\sqrt{6}}\lambda x_{ACC}&=&0.
\eea 
We can define an effective potential parameter, $\lambda_{eff}$,  that would give rise to the same dynamics in the absence of the quartic coupling, 
\bea
 \lambda_{eff}=\sqrt{6}x_{ACC}(F_q)
\eea
The Einstein frame effective equation of state
 \bea
 w_{E}
 &=& -1+\frac{1}{3}\lambda_{eff}^2(1 + 4 F_q\lambda_{eff}^2)
 \eea
As we see in Figure \ref{fig:lamQeff}, the effect of an increasing quartic coupling is to reduce the effective equation of state at late times relative to the standard ACC-$\lambda$ value of $w=-1+\lambda^2/3$. We can see this consistently in the analytic solutions in the limit of small $F_q$ for the effective equation of state in the Einstein frame 
\bea
w_E \approx -1 + \frac{\lambda^2}{3} - \frac{4 \lambda^4 F_q}{3}+...,
\eea
and in the Jordan frame
\bea
w_J\approx-1 + \frac{\lambda(\lambda-2Q)}{3(1-Q\lambda)} \left(1- \frac{4 \lambda^2 F_q}{(1-Q \lambda)}+...\right).
\eea
This should allow a larger range of values for $\lambda$ to be consistent with observations for $F_q>0$.

%--------------------------------------------
\subsection{Coupling to the Einstein tensor}
\label{section-ET}
%--------------------------------------------

The presence of a direct coupling of the scalar to the Einstein tensor, with an amplitude parameterized by $F_c$ (\ref{Fcdef}), modifies the Friedmann, acceleration and fluid equations:
\bea
\Omega_m(a)&=&1-(1+18 F_c)x^2 -y^2 -z^2,
\eea
\bea
&&-\frac{2}{3} \frac{\dot{H}}{H^2} \left( 1+\frac{6 F_c}{1+6 F_c} (1+18F_c) x^2 \right) = 1+(1+18F_c)x^2 
\nonumber
\\ &&\hspace{1.5cm}-y^2+\frac{1}{3} z^2- 4 \frac{F_c}{1+6F_c} \sqrt{6} x   \left[ (Q \Omega_m + \lambda y^2)\right],\hspace{0.75cm}
\eea
and
\bea
 x' &=&-3x-(1+\frac{6F_c}{1+6F_c})x\frac{\dot{H}}{H^2} \nonumber
\\ &&+\frac{\sqrt{6}}{2}\frac{1}{(1+6 F_c)}\left[\lambda y^2 + Q\Omega_m(a)\right].
\eea
Where  $|F_c|<1/18$ if the coupling to the Einstein tensor is to be subdominant to the canonical kinetic term.  

While the new coupling does not introduce a new attractor, nor change the predictions for the RAD-$\lambda$ or the MAT-$\lambda$ attractor, it does modify the other two attractors from their nominal values, determined by the leading order terms.  

%==================================================
\begin{figure*}[!t]
\bc
\includegraphics[width=0.45\textwidth]{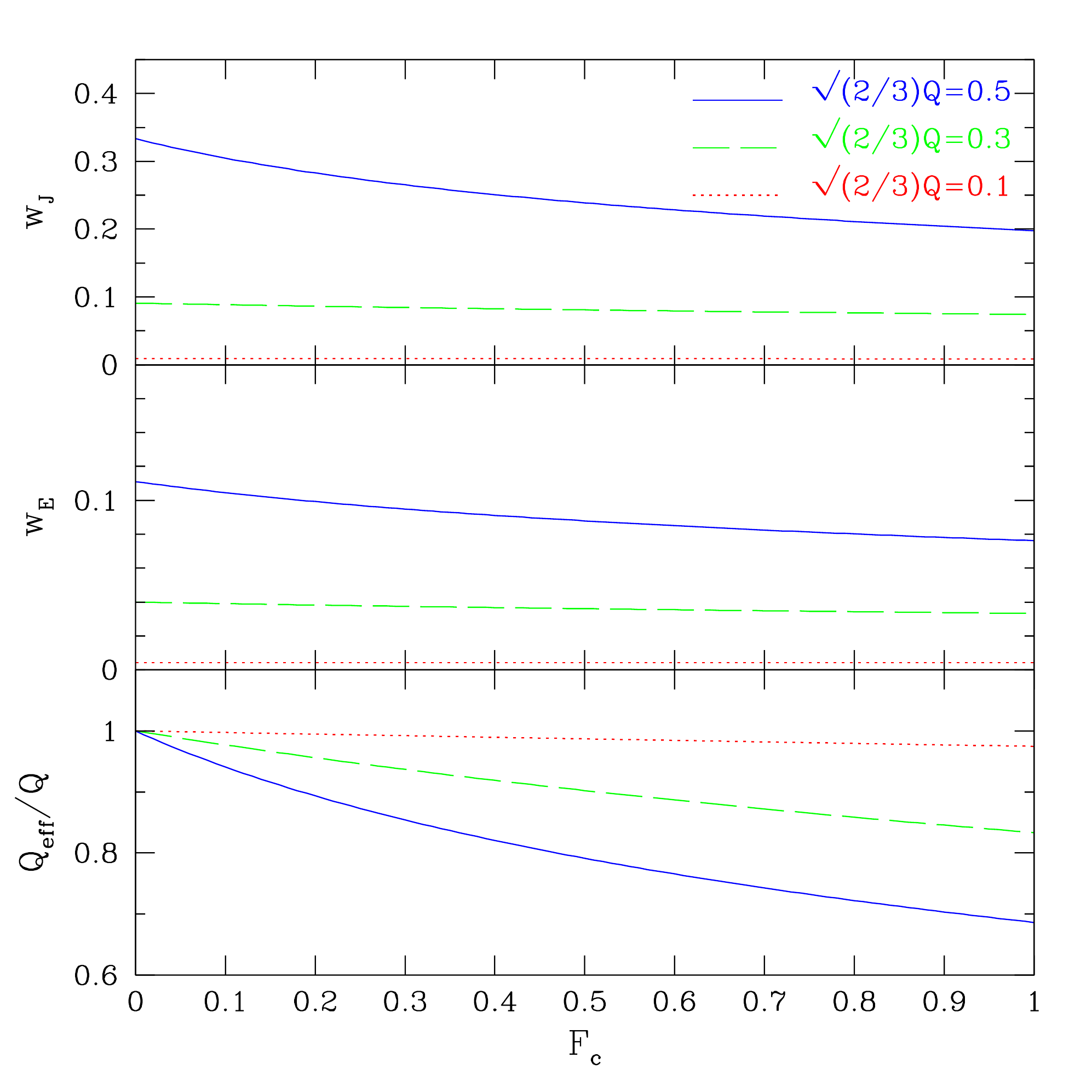}
\includegraphics[width=0.45\textwidth]{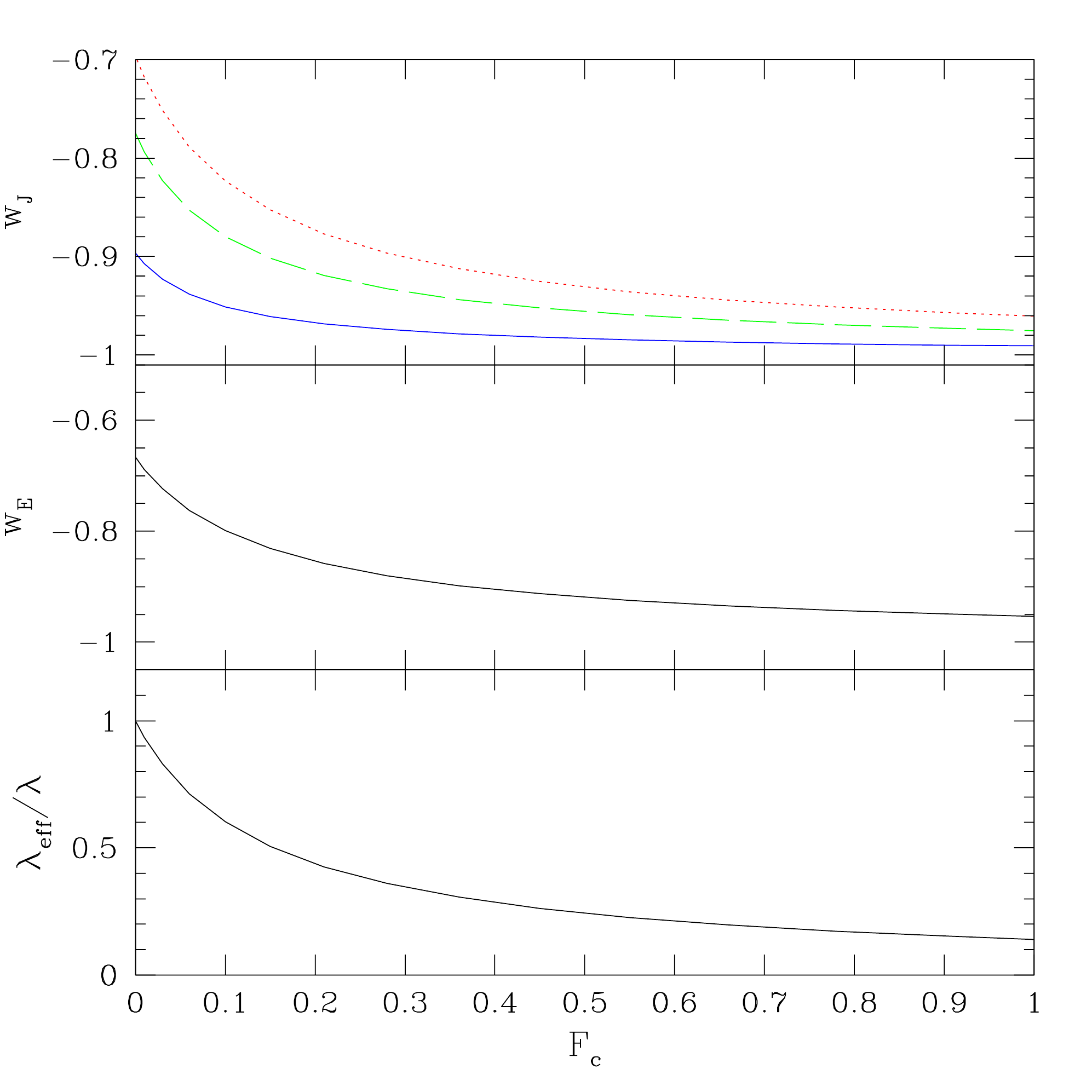}
\ec
\caption{The implications of the attractor solutions in the matter-dominated [left panel] and accelerative [right panel] eras when  the leading order terms in the action are augmented by a coupling to the Einstein tensor parameterized by an amplitude $F_c$. The panels show  [top] the effective equations of state in the Jordan frame,  $w_J$, and [middle]  the Einstein frame, $w_E$ along with [bottom] the effective coupling $Q_{eff}$ [right] and potential exponent $\lambda_{eff}$ [left], relative to their fiducial  values.  Three values of non-minimal couplings $\sqrt{3/2}Q=0.1,0.3,0.5$ are considered. During the accelerative era $Q$ only affects the conformal transformation and Jordan frame equation of state.}  \label{fig:curv}
\end{figure*}
%==================================================

The MAT-Q  solution now satisfies,
\bea
0 &=&\sqrt{6} Q - 3x - \sqrt{6}Qx^2(1+24 F_c) + (3+54 F_c)x^3 
\nonumber
\\ &&+6  \sqrt{6}(1+18F_c) F_c Q x^4
\eea
leading to the solution
\bea
x&=&\frac{1}{12 \sqrt{6} F_c Q} (-3 + 3\sqrt{1+16F_c Q^2}) 
\nonumber
\\ &\approx& \frac{2}{3}Q \left(1-4 F_cQ^2+...\right).
\eea
the approximation holding for $F_c\ll 1$.  The effective equation of state in the Einstein frame is reduced for $F_c>0$,
\bea
w_{E}&=& \frac{1}{12 F_c} \left( -1+ \sqrt{1+16F_c Q^2} \right)\nonumber
\\ & \approx& \frac{2}{3}Q^2\left(1-4F_c Q^2+...\right)
\eea 
as it also is in the Jordan frame,
\bea
w_J&=& \frac{-1+4Q^2+\sqrt{1+16Fc Q^2}}{3(1+2Fc-2Q^2)}
\\
&\approx& \frac{4 Q^2}{3(1-2Q^2)}\left( 1+ \frac{4 Q^2 F_c}{1-2 Q^2} \right).
\eea

 In the $MAT-\lambda$ solution, while the relative amplitudes of the kinetic and potential scalar densities are changed, 
\bea
x&=&\sqrt{\frac{3}{2}}\frac{1}{ (\lambda-Q)}
\\
y&=& \frac{\sqrt{\frac{3}{2}-Q(\lambda-Q)\left(1+\frac{9F_c}{\lambda-Q}\right)}}{\lambda-Q},
\eea
the effective equation of state is unchanged 
\bea
w_{E}&=&\frac{Q}{\lambda-Q}.
\eea

The late time accelerative attractor equation, with $y^2=1-(1+18)x^2$ and $z=0$, is
\bea
x^\prime&=&-3x+\frac{\sqrt{6}}{2} \frac{\lambda}{1+6Fc}  -3 \sqrt{6} \frac{F_c}{1+6F_c} \lambda x^2
\eea
altering the ACC-$\lambda$ solution
\bea
x&=&\frac{-1-6 F_c+\sqrt{1+4F_c(3+9Fc+\lambda^2)}}{2 \sqrt{6} F_c \lambda}
\\
&\approx& \frac{\lambda}{\sqrt{6}} \left[ 1-6F_c \lambda\left(1+\frac{\lambda^2}{6} \right) \right]+....
\eea
For small $F_c$ the effective accelerative equation of state is seen to become more negative for $F_c>0$,
\bea
w_{E} &\approx &-1+\frac{\lambda^2}{3}\left[1-6F_c\left(1+\frac{\lambda^2}{6}\right)\right]+...,
\eea
and
\bea
w_J &\approx & -1 + \frac{\lambda(\lambda-2Q)}{3(1-Q\lambda)}\left(1-F_c\frac{(\lambda^2+6)}{(1-Q\lambda)}\right)+....\hspace{0.25cm}
\eea
The numerical solutions are shown in Fig.\ref{fig:curv}. For increasingly positive values of $F_c$ the matter and accelerative era equations of state are lower, making them more consistent with data than $F_c=0$. While we show the implications for $0\leq F_c\leq 1$ here, in our analysis to follow we will impose a restriction that $|F_c|<1/18$ to ensure the term remains subdominant to those at leading order.

%--------------------------------------------
\subsection{A kinetic non-minimal coupling}
\label{sec-kincoup}
%--------------------------------------------
If we introduce a kinetic non-minimal coupling of the scalar field to matter, as in (\ref{Fkdef}), the Friedman equation and acceleration equation in the Einstein frame are unchanged, however the scalar equation of motion is modified leading to a modified attractor equation 
\bea
&& x'+ x \frac{\dot{H}}{H^2} =   \frac{1 - 6 F_{k} x^2}{1-6 F_{k} x^2 - 3 F_{k} \Omega_m} 
\\  &&\hspace{0.75cm}\times\left(-3 x + \sqrt{\frac{3}{2}} \lambda y^2 + \sqrt{\frac{3}{2}} \Omega_m Q - \frac{3 F_{kin} \Omega_m x}{(1-6 F_{k} x^2)} \frac{\dot{H}}{H^2} \right). \nonumber
\eea
This does not influence the effective equation of state in either the RAD-$\lambda$ or  the RAD-null era,  and the leading order matter and accelerative era attractors are unmodified by the inclusion of $f_{kin}$. Moreover, the conformal transformation  can still be calculated via (\ref{eq:weffJ}) since  only derivatives of  $f_{kin} \dot\phi^2 = 6 F_k x^2$  enter the equation which are zero for attractor solutions.  The coupling does, however open up additional matter and accelerative era attractor solutions, with $x=1/\sqrt{6F_k}$.
While we  find that the numerical analyses do in some circumstances attempt to approach this attractor, the effect tends to be a transitory,  then returning to the leading order attractors. In its limit this attractor would lead to an ill-defined and unphysical $\Omega=0$ in the conformal transformation to the Jordan frame. Since the well-defined attractors are  unmodified by this first order term, we do not investigate it further in the remainder of the paper. 

%%%%%%%%%%%%%%%%%%%%%%%%%%%%%%%%%%%%%%%%%%%%%%
\section{Comparison with Data}
\label{sec-constraints}
%%%%%%%%%%%%%%%%%%%%%%%%%%%%%%%%%%%%%%%%%%%%%

%%%%%%%%%%%%%%%%%%%%%%%%%%%%%%%%%%%%%%%%%%
\subsection{Analysis Approach}
\label{sec-mcmc}
%%%%%%%%%%%%%%%%%%%%%%%%%%%%%%%%%%%%%%%%%%

To investigate the impacts of the leading order and first order terms in the EFT, we numerically evolve the Einstein frame Friedmann, acceleration and scalar field fluid equations, and simultaneously  use the conformal transformation to calculate the Jordan frame variables. We compare the data to quantities in the Jordan frame since observations such as fluctuations in the CMB and the redshift of supernovae are reported  assuming that the baryons are minimally coupled to gravity.  We hence define the present day epoch as $a_J=1$ and using the Jordan frame redshifts and distance measures as the physical observables.

To establish the cosmological constraints on  the EFT parameters,  we perform a Monte Carlo Marcov Chain (MCMC) analysis assuming flat priors on  $\Omega_b$, $\Omega_m$ and $H_0$, along with leading order parameters $V_0$, $\lambda$ and $Q$.  We consider constraints on the leading order terms plus each first order term separately to understand the individual effects of each.   We assume flat priors on the quartic coupling, $-10\leq F_q\leq 10$, and the coupling to $G_{\mu\nu}$, $0\leq F_c\leq 1/18$ where the upper limit ensures sub dominance to the leading order terms. Scenarios with a Gauss Bonnet term are investigated with exponent $0\leq \mu\leq 70$ and magnitude given by $F_0 = p_{GB}F_{0}^{est}$ where $F_0^{set}$ was defined in (\ref{F0est})  and $-10 \leq log \ p_{GB}\leq 0$.

Our 1D and 2D constraints are obtained after marginalizing over the remaining parameters using the programs included in the publicly available  {\it CosmoMC} package \footnote{http://cosmologist.info}. To ensure convergence we apply the Gelman and Rubin Òvariance of chain meanÓ/Òmean of chain variancesÓ R statistic for each parameter on 8 or more chains.  All MCMC runs have a convergence of $R<0.1$ or lower.

%SN
We consider constraints from the ``Union 2.1"  compilation \footnote{http://supernova.lbl.gov/Union/} of 580 Type Ia supernovae observations \cite{Suzuki:2011hu}, with  redshifts in the range $0 <z<1.414$. 
We compare predicted distance modulus estimates for the MCMC scenarios  for each supernovae, at redshift z,
\be
\mu(z) = 5 \; log\left[ \frac{D_L(z) }{1Mpc} \right]+ 25,
\ee
where $D_L$ is the luminosity  distance,  against the observations using the compilation's covariance matrix including systematic errors.

%CMB
To investigate geometric constraints from the CMB, we  use the WMAP-7 \cite{Komatsu:2010fb}  results. The CMB is sensitive to two distance ratios to decoupling, through the position of the peaks and the acoustic oscillations \cite{Komatsu:2008hk}:  the acoustic scale at decoupling,
\bea
l_A(z_*) &\equiv  (1+z_*)\frac{\pi D_A(z_*)}{r_s(z_*)},
\eea
and the `shift parameter', 
\bea
\mathcal{R} &\equiv \frac{\sqrt{\Omega_m H_0^2}}{c} (1+z_*) D_A(z_*),
\eea
where $D_A(z_*)$ is the angular diameter distance, and $r_s(z_*)$ is the sound horizon, at  the redshift to decoupling $z_*$.  

As discussed in \cite{Komatsu:2008hk}, the definition of $\mathcal{R}$ suppresses the influence of radiation, dark energy or curvature on the Hubble parameter at decoupling $H(z_*)$ but is used by convention. We use the CMB data vector given in \cite{Komatsu:2010fb}  WMAP 7 results 
\[
	\begin{pmatrix}	
	l_A                \\ 
	\mathcal{R}  \\
	z_*             
	\end{pmatrix}
	=  \begin{pmatrix}
		302.09 \\
		1.725 \\
		1091.3
              \end{pmatrix}				
\]
with the inverse covariance matrix
\[
 Cov^{-1}_{CMB} = 
\begin{pmatrix}
2.305 & 29.698     & -1.333      \\
           & 6825.270 & -113.180 \\
	  &                    &3.414 
\end{pmatrix}.
\]
The redshift of decoupling, $z_*$, is obtained , to percent accuracy, from the fitting formula \cite{Hu:1995en}
\begin{align}
g_2 &=\frac{0.560}{1+21.1 (\Omega_{b} h^2)^{1.81}}, \\
g_1 &= \frac{0.0783 ( \Omega_m h^2)^{-0.238}}{1+39.5 (\Omega_b h^2)^{0.763}}, \\
z_* &= 1048 (1+0.00124 (\Omega_b h^2)^{-0.738}) ( 1 + g_1 (\Omega_m h^2)^{g_2}).
\end{align}
where $h=H_0/100kms^{-1}Mpc^{-1}$. In principle a non-minimal coupling, $Q$, will affect the  the matter dominated expansion era and consequently the accuracy of the fitting function. We find that the redshift to decoupling $z_*$ is only changed by $2 \times 10^{-3} $ \%, and  the effects on $l_A$ and $\mathcal{R}$ are smaller, so that the effect of the coupling on the accuracy fitting function in not a significant concern. 

%BAO
Baryon Acoustic Oscillations (BAO) are the imprint of the sound horizon at last scattering as a characteristic scale  in the clustering of matter.  
When observed at different redshifts the characteristic scale can be used as a  Standard Ruler to estimate cosmological distances. 

The acoustic scale along the line of sight encodes information about the Hubble parameter $H$ whereas the tangential component measures the angular diameter distance $D_A$.  Current measurement accuracy  is not  sufficient to measure $H_0$ and $D_A$  separately, so observational radial and tangential measurements are typically combined into an effective, averaged scale \cite{2005ApJ...633..560E}  defined as
\begin{equation}
D_v(z)= \left[\frac{D_A(z)^2(1+z)^2 c z}{H(z)} \right]^{1/3}.
\end{equation}
The most accurate measurements of the BAO results to date come from 2dFGRS, SDSS DR7, WiggleZ and BOSS spectroscopic redshift surveys.
The results presented in \cite{Percival:2007yw} are based on the spectroscopic SDSS DR7 sample, including both LRG and Main galaxy samples in combination with the 2dFGRS survey. The ratio $r_s(z_{drag})/D_v(z)$ is given for two redshifts, $z=0.2$ and $z=0.35$,
\bea
r_s(z_{drag})/D_v(0.35)=0.109715,  \\
r_s(z_{drag})/D_v(0.2)=0.190533
\eea
 with the inverse covariance matrix 
\[
 Cov^{-1}_{BAO:SDSS} = 
\begin{pmatrix}
30124 & -17227        \\
            & 86977
\end{pmatrix}.
\]
 $z_{drag}$ is the comoving sound horizon at the baryon drag epoch, baryon decouple from photons.  If the ratio $3 \rho_b/ 4 \rho_\gamma=1$ at $z_*$ then the drag epoch and decoupling occur simultaneously. For typical, cosmological scenarios, however $z_*> z_{drag}$. We calculate the redshift $z_{drag}$  using the fitting formula from \cite{ Eisenstein:1997ik}
\bea
z_d& =& 1291 \frac{(\Omega_0 h_0^2)^{0.251}}{1 + 0.659 (\Omega_0h_0^2)^{0.828}} [1 + b_1 (\Omega_b h_0^2)^b2], 
\\
b_1 &=& 0.313 (\Omega_0 h_0^2)^{0.419} [1 + 0.607 (\Omega_0 h_0^2)^{0.674}],
\\
b_2 &=& 0.238 (\Omega_0 h_0^2)^{0.223},
\eea
accurate to a few percent.
 
As the distance scale $D_v$ is highly degenerate with $\Omega_{m}h^2$ the WiggleZ survey introduced the acoustic parameter $A(z)$ \cite{Blake:2011en},
\begin{equation}
A(z)=D_v(z) \sqrt{\Omega_mH_0^2}/cz.
\end{equation}
The WiggleZ survey provides BAO measurements for three redshifts complementary to those from the SDSS/2dFGRS surveys: $A(0.44)=0.474$, $A(0.6)=0.442$, $A(0.73)=0.424$.
The inverse covariance matrix is given by
\[
 Cov^{-1}_{BAO:WiggleZ} = 
\begin{pmatrix}
1040.3 & -807.5     & 336.8      \\
           & 3720.3 &  -1551.9 \\
	  &                    &2914.9
\end{pmatrix}.
\]
Finally, we include the recent Baryon Oscillation Spectroscopic Survey (BOSS) survey results, $D_v/r_s=13.67\pm0.22$ at $z=0.57$ \cite{Anderson:2012sa}. 

 In combination, the data sets  have $589-d$ degrees of freedom where $d$ is the number of MCMC parameters. For $\Lambda CDM$ $d=3$, leading order $d=6$ and all other model have $d=7$.

%--------------------------------------------
\subsection{Findings}
\label{sec-findings}
%--------------------------------------------

%--------------------------------------------
\subsubsection{Constraints on the leading order,  quartic kinetic and  Einstein tensor terms}
\label{sec-mcmc_lead}
%--------------------------------------------

In Table \ref{tab:mcmc}  we summarize the results of the MCMC analysis for the EFT models  involving only the leading order terms, and those in which a quartic coupling or a coupling to the Einstein tensor are present, in comparison to $\Lambda$CDM. These models modify the attractor behaviors in both the matter and accelerative eras. The minimum $\chi^2$  is the same for each scenario, $\chi^2=547.0$,  equivalent to that for $\Lambda$CDM,  reflecting that in spite of including one or two extra parameters, these scenarios can recreate, but not improve upon, the predictions for $\Lambda$CDM. 

%==================================================
\begin{table}[!t]
\bc

	\begin{tabular}{|l |c |c |c |}
		\hline
			    Model                           & $\Omega_{m}$ & $ |Q|$ &  $|\lambda|$   \\ \hline \hline
		             $\Lambda$CDM         &    $0.291_{-0.028}^{+0.031}$    &   ---                          &   ---                                     \\ \hline 
	                       +Q (leading order)     &    $0.291_{-0.030}^{+0.033}$ &   $<0.043$     &   $<1.15$     \\ \hline 
	                       +$F_{c}$	               &	$0.292_{-0.029}^{+0.033}$ &   $<0.043$       &    $<1.40$ 	\\ \hline
			    +$F_{q}$               &    $0.293_{-0.030}^{+0.033}$     &   $<0.044$    &   $<2.00$      \\ \hline 
		\end{tabular}

\ec
	\caption{Summary of the 95\% confidence level constraints from the MCMC analysis for all scenarios  except that including the Gauss-Bonnet term. The $\chi^2$ for all models are the same as for $\Lambda CDM$= $547.0$
	}		
	\label{tab:mcmc}
\end{table}
%=====================================================

%==================================================
\begin{figure*}[!t]
\bc
\includegraphics[width=0.4\textwidth]{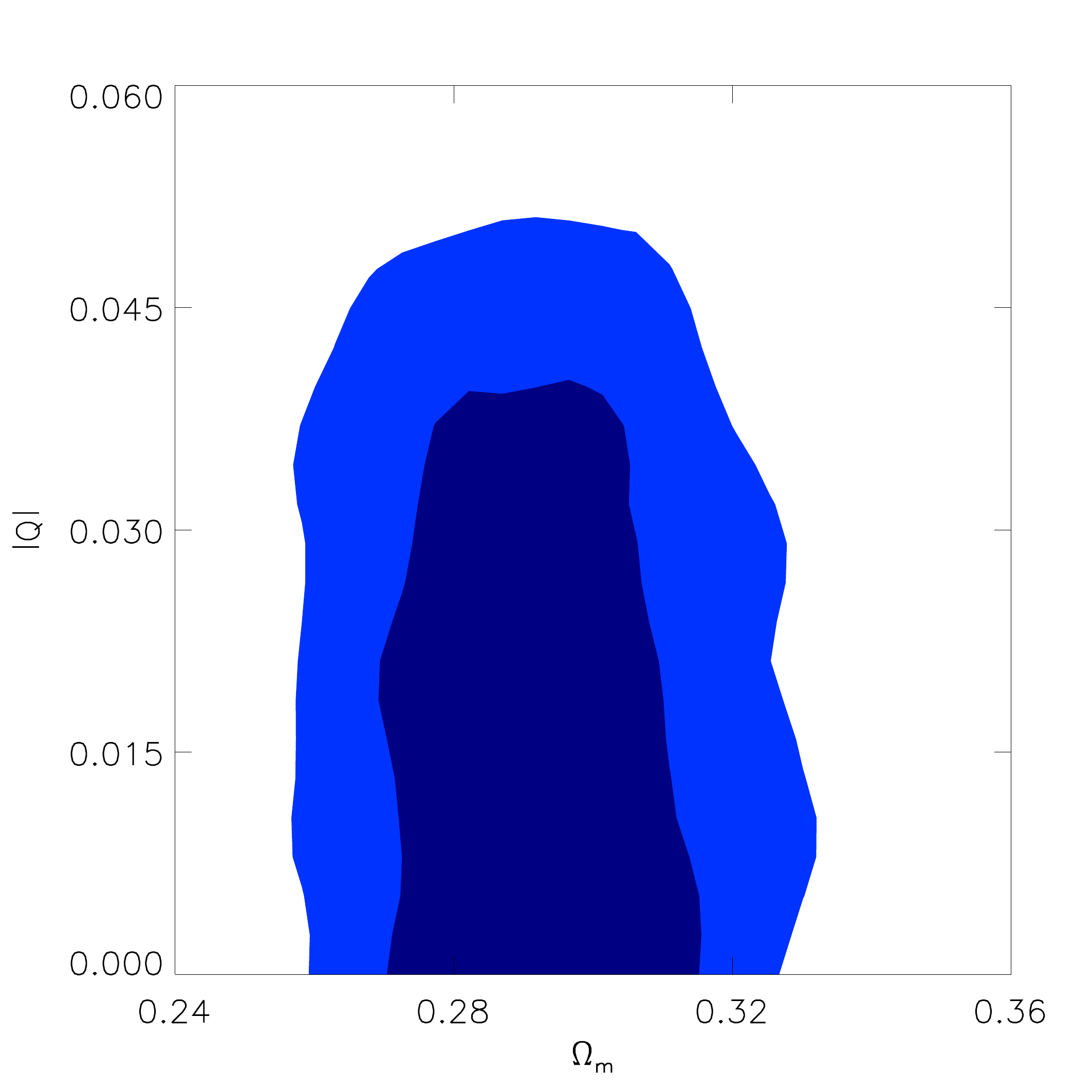}
\includegraphics[width=0.4\textwidth]{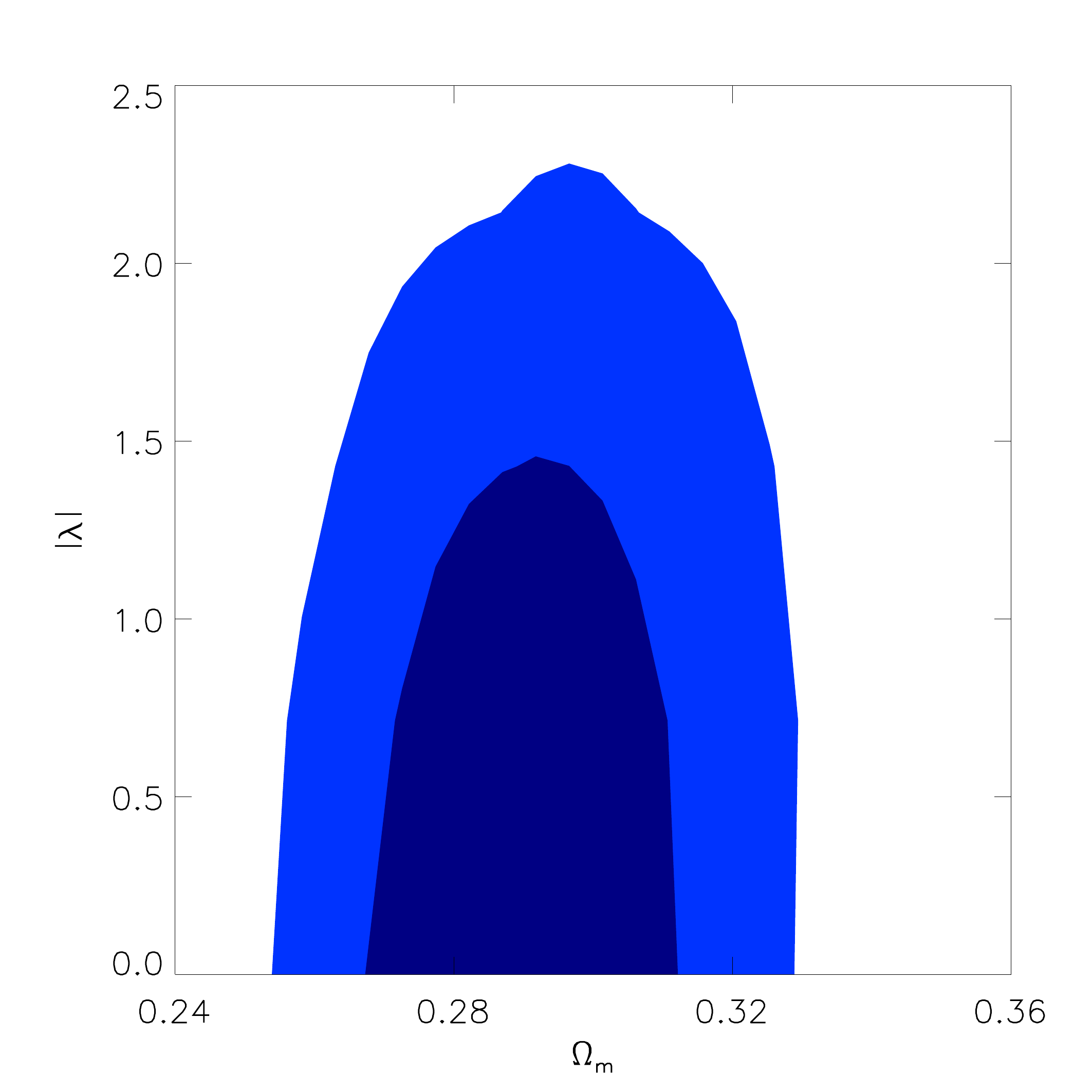}
\ec
\caption{ Joint $68\%$ (dark shaded) and $95\%$ (light shaded) constraints for the leading order action  on the fractional matter density today, $\Omega_m$ and the coupling of the scalar field to matter, $Q$, [left panel]  and the scalar potential exponent, $\lambda$ [right panel].  }\label{fig:leading}
\end{figure*}
%==================================================

%==================================================
\begin{figure*}[!t]
\bc
\includegraphics[width=0.4\textwidth]{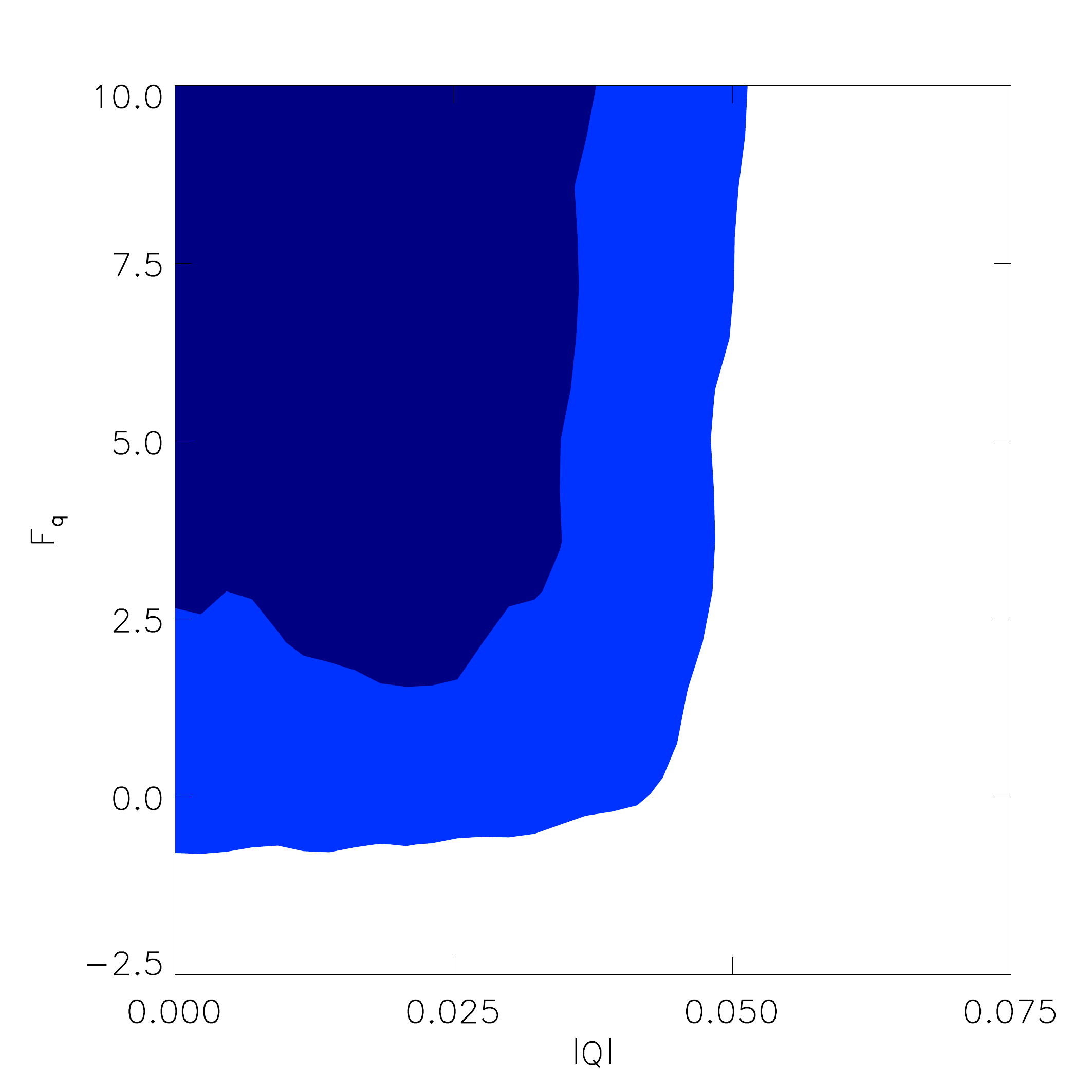}
\includegraphics[width=0.4\textwidth]{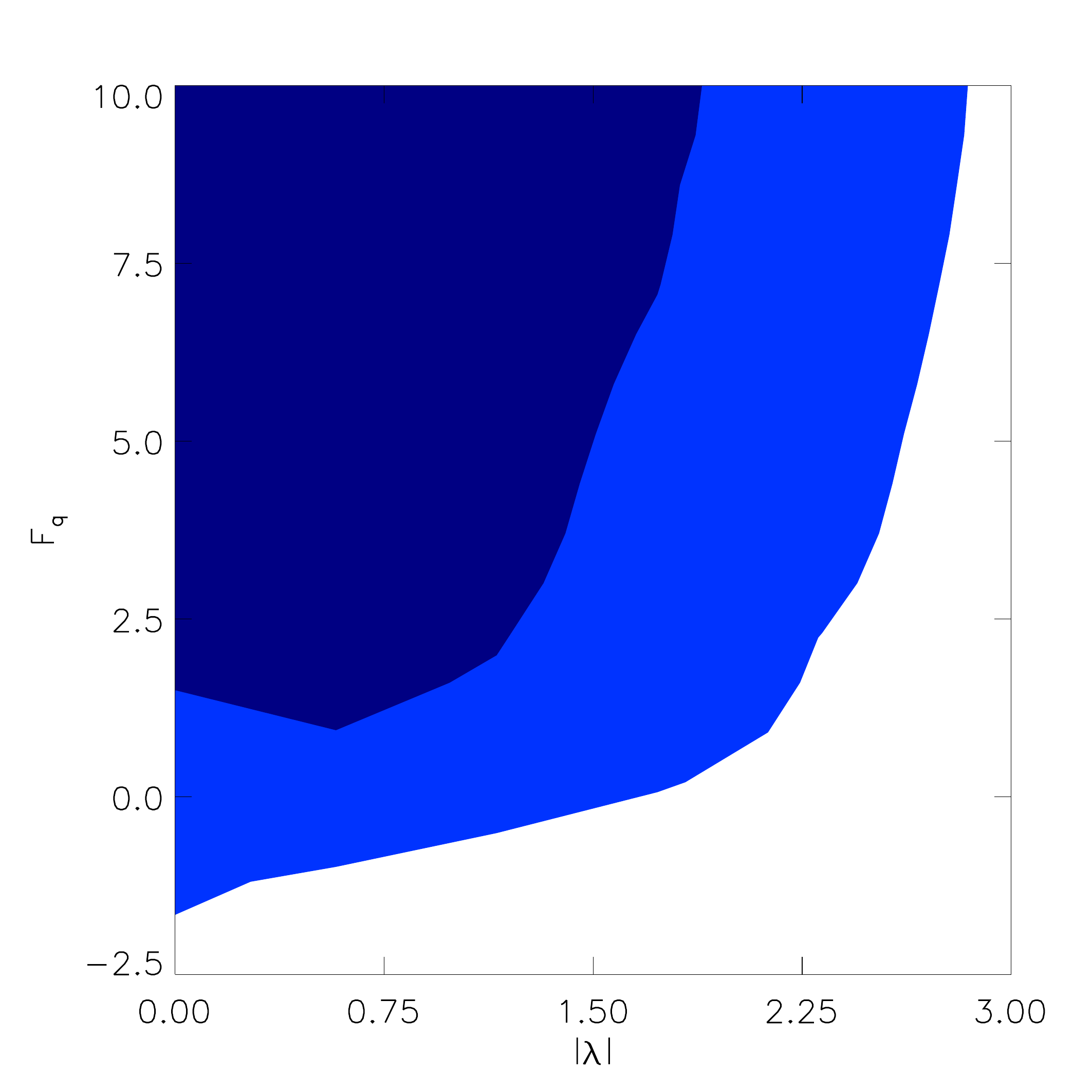}

\ec
\caption{Joint $68\%$ (dark shaded) and $95\%$ (light shaded) constraints on the quartic kinetic coupling, $F_q$ and the coupling of the scalar field to matter, $Q$, [left panel]  and the scalar potential exponent, $\lambda$   [right panel]. The quartic term has a significant effect on reducing the equation of state in the accelerative era, allowing a larger range of $\lambda$ to be consistent with the data. The evolution is sensitive to the sign of $F_q$ leading to a lower bound on the coupling.}\label{fig:Fq}
\end{figure*}
%==================================================

Scenarios with increasing magnitudes of coupling and exponential exponent  come into tension with the data as they increase the predicted equation of state parameter in  the matter era (for $Q\ne0$) and the accelerative era (for $\lambda\ne0$)  relative to the $\Lambda$CDM prediction. In the minimally coupled scenario, one would require $\lambda<\sqrt{2}$ to achieve any acceleration ( and $\lambda\ll \sqrt{2}$ to have $w\approx-1$).  The presence of a non-minimal coupling, $Q< 1/\lambda$, creates a more negative equation of state in the Jordan frame, allowing a larger range of $\lambda$, including potentially $\lambda>\sqrt{2}$, to be consistent with the data. 

%=================================================================
\begin{figure}[!h]
\bc
\includegraphics[width=0.5\textwidth]{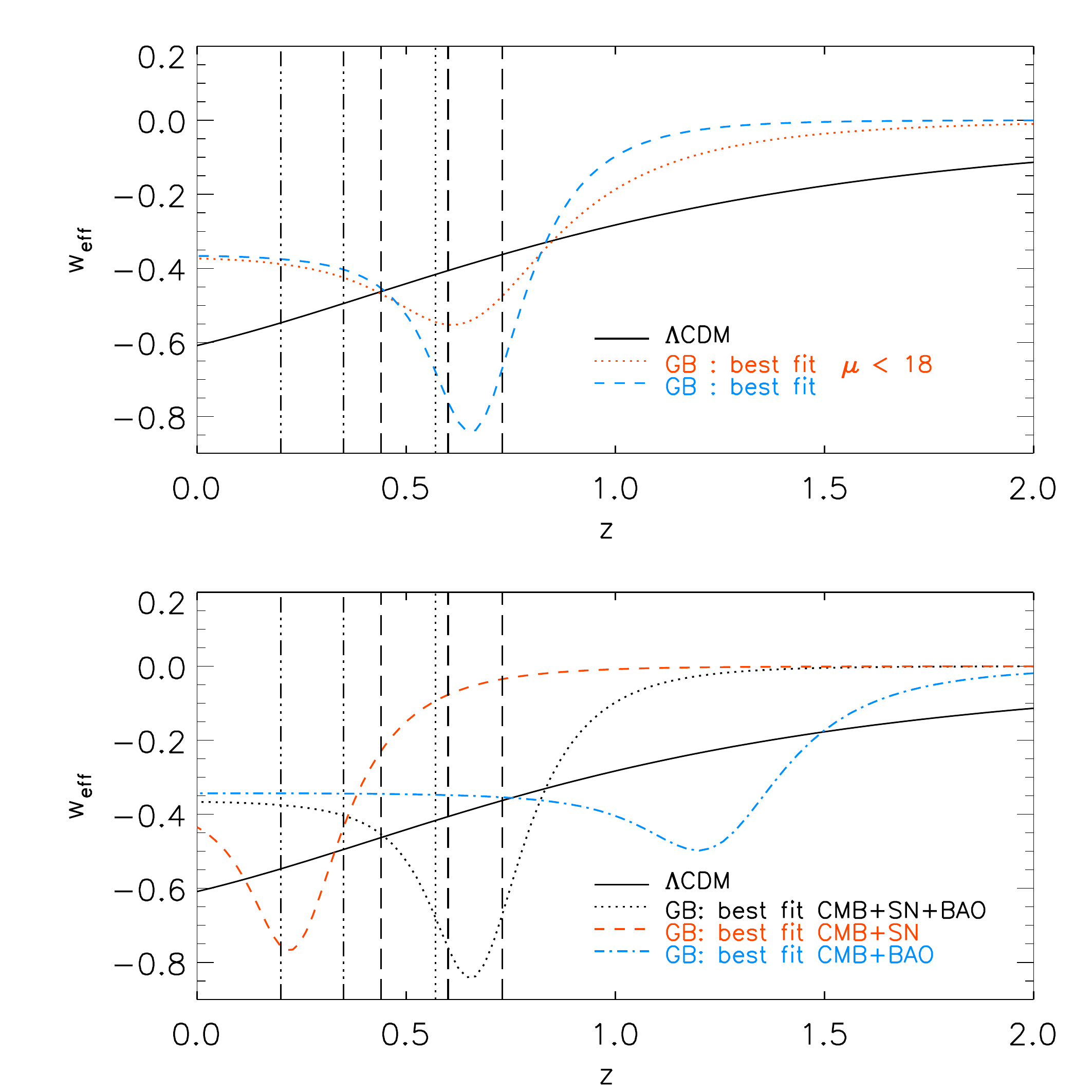}

\caption{The effective equation of state, $w_{eff}$,  as a function of redshift, $z$,for the  Gauss Bonnet model  in comparison to a best fit  $\Lambda CDM$ scenario (black solid). [Top panel] Comparison of the best fit prediction for the Gauss-Bonnet model overall [blue,dashed] and for models with a low value of the exponent, $\mu<18$ (red dotted), showing how the depth of the ``trough" in $w_{eff}$ is  dependent on $\mu$. [Lower panel]   Comparison of the overall best fit for CMB+SN (blue, dot- dashed), CMB+BAO (red, dashed dotted) and CMB+BAO+SN (black, dotted) showing the tension between the preferred evolution histories for the different datasets.  } \label{fig:GB_weff}
\ec
\end{figure}
%==================================================================

In Figure \ref{fig:leading}  we show the combined 2D constraints arising from CMB, BAO and SN for the coupling Q and the exponential exponent $\lambda$.   We find 1D marginalized errors on the coupling of $|Q| < 0.026 (0.043)$ and exponential potential, $\lambda<0.80 (1.15)$ at the $68\%$ ($95\%$) confidence level.    In \cite{Bean:2008ac} a similar analysis was performed, but in the context of constraints arising from a non-minimal coupling purely to cold dark matter, for which the comparison with observations is performed in the Einstein frame.  They found similar constraints of $|Q| <  0.055$ and exponential potential, $\lambda<0.95$ at the $95\%$ confidence level. The similarity can be understood in terms of the small values of $Q$; as the conformal transformation tends towards unity, the Einstein and Jordan frame become comparable.

As discussed in section \ref{sec-attractor}, while the quartic kinetic term and the coupling to the Einstein tensor don't lead to new attractor solutions, they do alter the leading order attractor solutions during the matter dominated and accelerative eras.   Their effects become less significant, however, as the coupling, $Q$, becomes smaller. We find  that the constraints on $Q$ are little changed by the inclusion of these terms as its magnitude is already tightly constrained, to be very small, by the data. During the accelerative era these terms have a more pronounced effect, and their enhancement  of the accelerative equation of state  allows a broader range of $\lambda$ to be consistent with the data, as given in Table \ref{tab:mcmc}. 

While the effects of $Q$ and $\lambda$ are sensitive to their magnitude, not sign,  the quartic coupling term modification  is sign dependent; negative values of $F_q$ and $F_c$ increase the effective equation of state during both the matter and accelerative eras. If the quartic term is included, we find the data provide a lower bound with $F_q>0.22 $ at the 95\% confidence level, as shown in Figure \ref{fig:Fq}.   For the coupling to the Einstein tensor, we find no significant difference in best fit likelihoods in the range we investigated $0\leq F_c\leq 1/18$.

%=================================================================
\begin{figure}[!t]
\bc
\includegraphics[width=.5\textwidth]{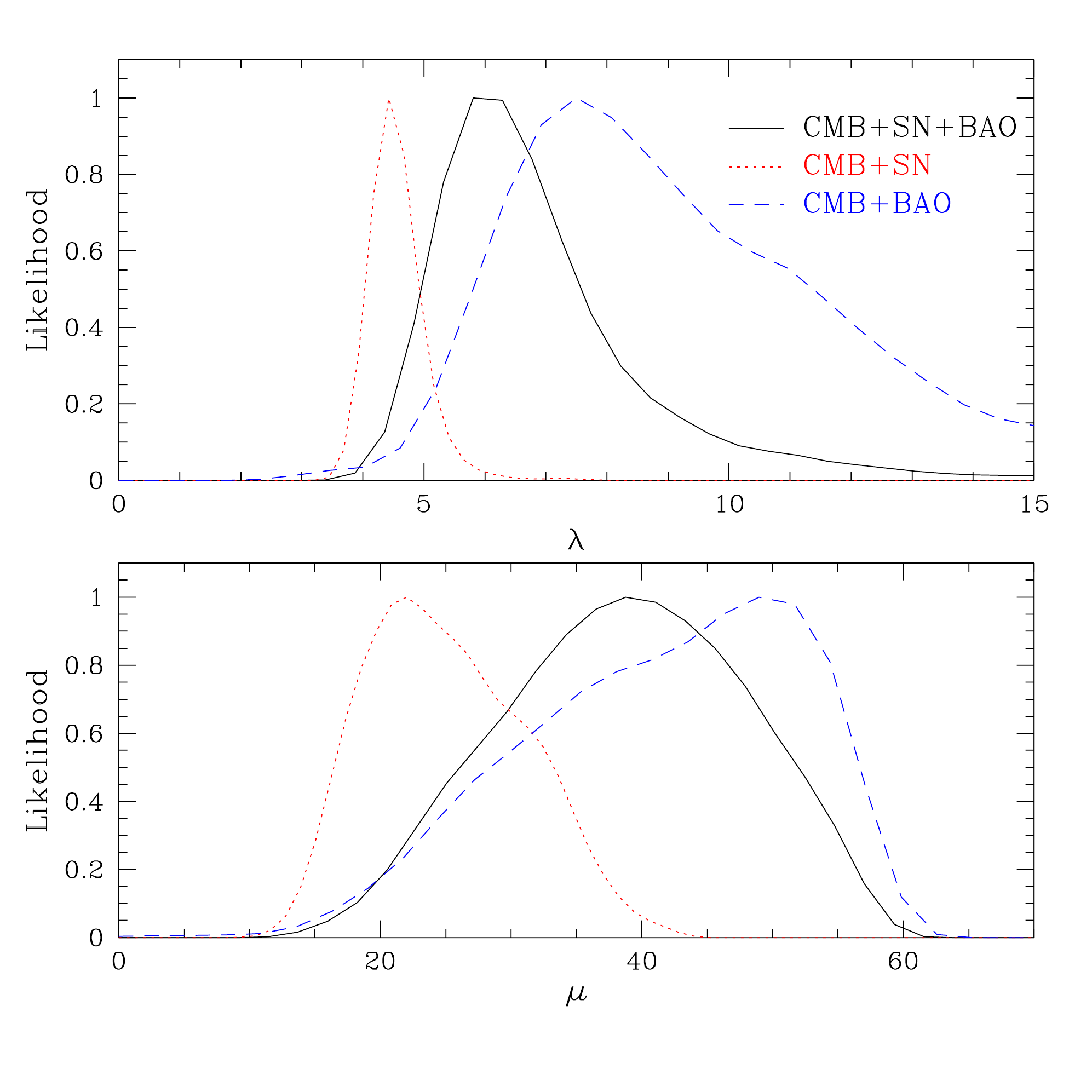}

\caption{ 1D likelihood contours for the Gauss-Bonnet parameter $\mu$ and the  potential exponent $\lambda$ for the different data sets:  CMB+BAO+SN (black solid line),  CMB+BAO (blue dashed line) and CMB+SN (red dotted line ). There is  a clear tension between the supernova and the BAO data sets. As discussed in the text, through their effects on $w_{eff}$, the data provide both upper and lower bounds on $\lambda$ and $\mu$ in this model. } \label{fig:GB_1D}
\ec
\end{figure}
%==================================================================

%=================================================================
\begin{figure}[!t]
\bc
\includegraphics[width=0.5\textwidth]{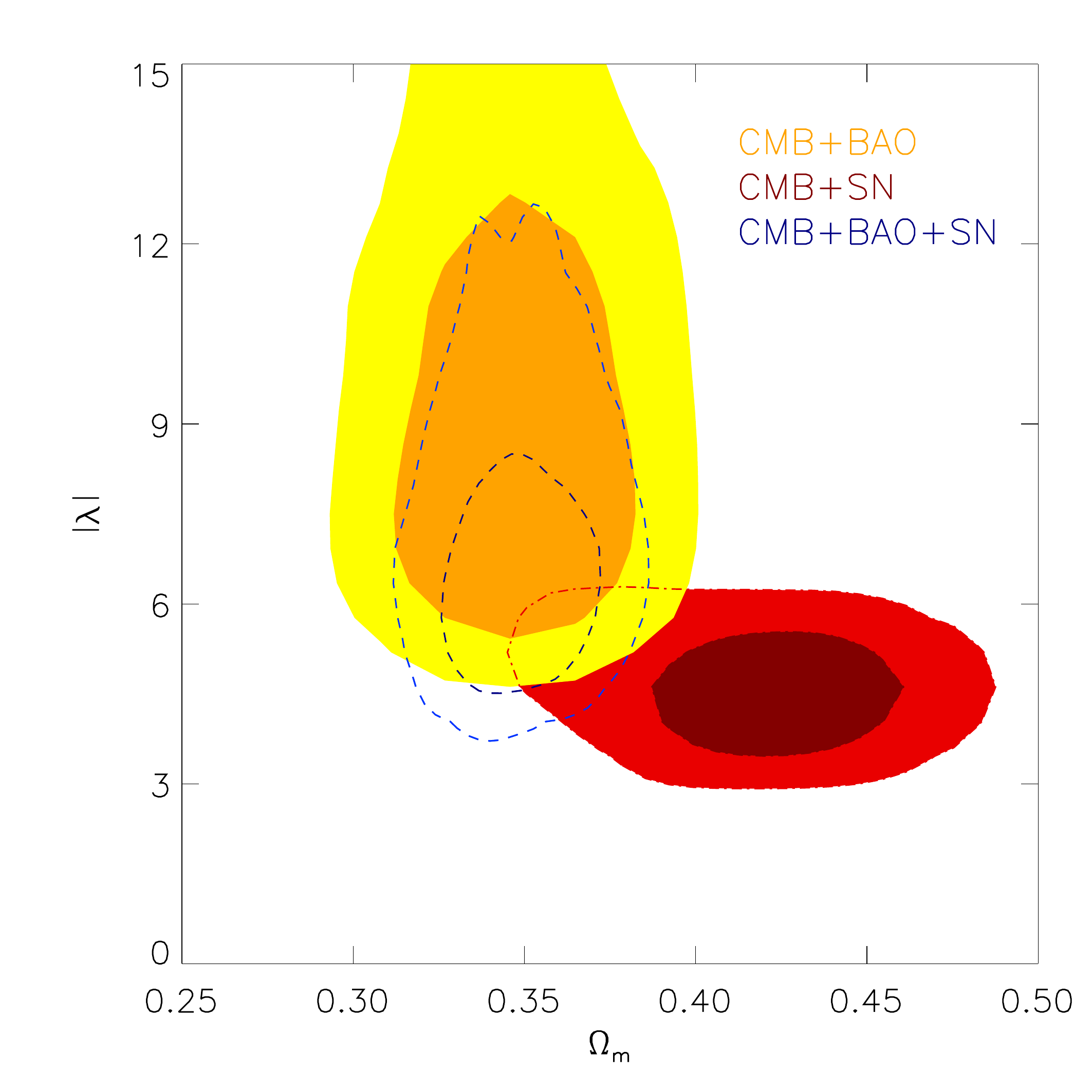}

\caption{ Joint $68\%$ (dark shaded) and $95\%$ (light shaded) constraints for the fraction matter density, $\Omega_m$ and for potential exponent $\lambda$ in the Gauss-Bonnet model for the different data sets:  CMB+SN (red solid contours),  CMB+BAO (yellow solid contours), and in combination CMB+BAO+SN (blue dashed lines). There CMB+SN and CMB+BAO data have almost orthogonal dependencies in this projection, highlighting the origins of the tension between them. } \label{fig:GB_2D}
\ec
\end{figure}
%==================================================================

%--------------------------------------------
\subsubsection{Constraints on the Gauss-Bonnet term}
\label{sec-mcmc_GB}
%--------------------------------------------
The Gauss-Bonnet term  gives rise to a new accelerative attractor solution through creating a minimum in the effective scalar potential, produced by the exponential potential and the GB term. This enables the expansion history to exit the scaling solution during the matter dominated era into a stable de-Sitter point.   The GB term does not affect the RAD or MAT era attractors, so we consider its impact on a simple non-minimally coupled quintessence model, with $Q=0$.

 In this scenario our present day epoch is in the transition period between the matter and de-Sitter eras characterized by a trough \cite{Tsujikawa:2006ph} and plateau in the   effective equation of state as shown in Figure \ref{fig:GB_weff}.  Prior to the transition the scalar follows the MAT-$\lambda$ attractor with $w_{eff}=0$. At the transition epoch the positive gradient of GB term temporarily dominates the field's equation of motion.  The trough in the effective equation of state is  generated as the scalar slows and instantaneously becomes static, then $\dot\phi$ changes sign, and the equation of state increases again. Following this the scalar proceeds on a slow evolution towards the static de-Sitter solution (in the distant future).  Both before and after the trough, $w_{eff}$ is greater than in $\Lambda$CDM scenarios. 
 
 We note that wfhile in the matter and de-Sitter eras the fractional energy density in the GB term, $\Omega_{GB}=\sqrt{6}xv$, tends to zero, it is non-zero and can be significant in this brief transition between the two. During this epoch therefore we are close to, if not at the point, of no longer satisfying the sub dominance criteria used  to formulate the EFT action.

The depth of the trough is enhanced by a smaller value of $\lambda$ (giving a larger kinetic energy during the MAT-$\lambda$ phase), or steeper GB coupling, $\mu$.  Enhancing the depth of trough also  reduces the duration of the transient feature, however, so that there is a sweet spot optimizing the trough's observational impact. The redshift position of the transient is degenerately sensitive to $\mu$, $\lambda$ and $V_0$ through (\ref{F0est}) and can be tuned through varying $p_{GB}$.

 As summarized in Table \ref{tab:mcmc_GB}, we find that this distinctive equation of state profile is well constrained by the BAO and SN data, allowing us to place  constraints on $\mu$, $\lambda$ and $p_{GB}$, and  the contribution of the GB term in this effective action.  
 
In Figure \ref{fig:GB_1D} we show the tension between the constraints from CMB+SN and CMB+BAO in terms of the exponents $\mu$ and $\lambda$ and $\Omega_m$. We find the SN data have the strong preference for the GB trough to be located around $z\sim 0.2-0.3$, where previous  principal component analyses of SN  have shown the best measured modes in $w_{eff}$ are peaked \cite{Huterer:2002hy}.   The accelerative era starts somewhat later than in $\Lambda$CDM and corresponds to a significantly higher $\Omega_m$.  Values of $\lambda\sim 4-5$ are preferred to create a  trough that is deep but still sufficiently broad to make the  redshift averaged effective equation of state as consistent with that for $\Lambda$CDM as possible.  By contrast the BAO place tight constraints on the matter density today. The average equation of state in the low redshift regime is somewhat larger than that predicted in $\Lambda$CDM, which results in a comparatively higher $\Omega_m$ being preferred to fit the BAO data. This is consistent  with the standard degeneracies found in BAO constraints of a constant equation of state, e.g. \cite{Blake:2011en}. The BAO are less sensitive to shape and depth of the trough than the SN, and place weaker constraints on $\lambda$, $\mu$ and $p_{GB}$.

A previous analysis of this scenario \cite{Koivisto:2006ai}, only including the SDSS BAO survey, found reasonable agreement with the data when only considering CMB+BAO constraints. We find that when we also include the  WiggleZ and BOSS data the combined fit is not good and both the CMB+SN and CMB+BAO yield a worse fit than $\Lambda CDM$. The major issue arises when trying to  jointly fit SN and BAO data. In this case, the Gauss-Bonnet model fits the data significantly worse than $\Lambda CDM$, with a difference of $\chi^2$(GB)-$\chi^2(\Lambda$CDM)=17.  
%==================================================
\begin{table*}[htb]
\bc	
		\begin{tabular}{ |l || c |c|| c| c| c| c| c|}
		\hline
			\multicolumn{1}{|c||}{Data}  & \multicolumn{2}{|c||}{$\Lambda$CDM} & \multicolumn{5}{|c|}{Gauss-Bonnet}  \\ \hline
		                                      & $ \chi^2$	  & $\Omega_{m}$ & $\Delta \chi^2$ & $\Omega_{m}$ & $\lambda$ &  $\mu$ & lg $p_{GB}$ \\ \hline
			CMB+SN            &    545.3         &         $0.29_{-0.05}^{+0.05}$     &     2.8      &   $0.42_{-0.06}^{+0.05}$  &   $4.6_{-0.5}^{+0.7}$   &    $25_{-9}^{+10}$  & $-1.7^{+0.5}_{-0.7}$ \\ \hline 
			CMB+BAO         &    1.8              &         $0.29_{-0.03}^{+0.03}$     &     3.2      &    $0.35_{-0.03}^{+0.03}$   &   $9.0_{-3.1}^{+4.4}$   &    $41_{-18}^{+14}$&  $ -2.9^{+1.2}_{-1.1}$                \\ \hline 
			CMB+BAO+SN &    547.0          &        $ 0.29_{-0.03}^{+0.03}$    &     17.0    &    $0.35_{-0.03}^{+0.03}$   &   $6.8_{-1.8}^{+3.6}$   &    $39_{-15}^{+15}$    & $ -2.2 ^{+1.0}_{-1.3}$\\ \hline 
		\end{tabular}
		\caption{Summary of the $95\%$ confidence level constraints from Gauss-Bonnet MCMC analysis.  The SN data provide the best constraints on GB parameters $\mu$ and $\lambda$ and  $p_{GB}$. There is significant tension between the  constraints from SN and BAO datasets separately. The compromise, when the datasets are considered in combination, is a significantly worse fit to the data than $\Lambda CDM$. 	}
		\label{tab:mcmc_GB}
\ec
\end{table*}
%=====================================================

%%%%%%%%%%%%%%%%%%%%%%%%%%%%%%%%%%%%%%%%%%%%%%
\section{Conclusions }
\label{sec-conclusions}
%%%%%%%%%%%%%%%%%%%%%%%%%%%%%%%%%%%%%%%%%%%%%

We have considered the existence of dynamical attractor solutions in a general approach to dark energy model building, utilizing methods of effective field theory.  A wide range of dark energy and modified gravity models are able to be described using this approach, so that it forms a useful phenomenological link between underlying theories and observations. Dynamical attractors are powerful because of their weak sensitivity to initial conditions, which helps to alleviate some of the fine-tunings required to make models observationally viable. This robustness can also be useful in determining constraints on the action of the effective theory and in the presence of stark tensions with astrophysical observations, can help isolate terms which are strongly disfavored -- helping focus model building efforts.  

We considered both the analytical and numerical predictions  for the cosmic expansion history and obtained numerical constraints on the effective theory in light of recent CMB, Type Ia SN and BAO constraints. In the Einstein frame  attractor solutions exist that predict observationally consistent radiation (RAD-null) and accelerative eras (ACC-$\lambda$), but require tight constraints on  the presence of a non-minimal coupling to matter,$Q$, to give a viable matter era evolution (MAT-$Q$). We have shown that the addition of terms which are quartic in the time derivative of the scalar field, scalar couplings to the Einstein tensor, and a Gauss-Bonnet term can all lead to modifications of the expansion history in the matter and accelerative eras.  The quartic and Einstein tensor terms both modify the existing attractor solutions, creating effective couplings and potential exponents $Q_{eff}$ and $\lambda_{eff}$ that can reduce the effective equation of state parameter in both eras, and improve the fit to the data.  They have a limited impact on the constraints of $Q$ in the matter era,  because their effect is diminished when $Q$ is small. They can, however, have a significant affect on expanding the range of potentials consistent with the data by increasing the range of $\lambda$ allowed. 

The Gauss-Bonnet coupling opens up a new late-time de-Sitter solution (ACC-GB) induced by the creation of a minimum in the effective potential formed from $V(\phi)$ and the GB term. In this scenario, our current epoch  is a transitory era between  matter domination and the de-Sitter phase, with a characteristic evolution in the equation of state parameter that is constrained well by the data.
In particular, a tension exists between the BAO and Type Ia supernovae, through their different redshifts sensitivity, when fitting the GB model to the data. In combination they rule out this scenario at large significance, with a best fit $\chi^2(GB)-\chi^2(\Lambda$CDM)=17. This model, as well as being observationally disfavored, involves a transient epoch where the GB term becomes comparable with the leading order potential, so that on theoretical grounds one must also be cautious and ensure that the EFT remains valid.

The complementarity of the SN, BAO and CMB distance measures, has enabled us to place constraints on the cosmological background evolution by constraining the EFT couplings of higher dimensional operators. Our general approach is limited to perturbatively constructed backgrounds and does not apply to models where the background evolution becomes strongly coupled or highly non-linear.  This means that we are unable to capture screening effects that could be an important additional signature for some models of modified gravity, as well as models such as k-essence where a perturbative description of the background is not possible.  To utilize the EFT approach in these models,  one must put aside the possibility of constraining the background evolution, and instead construct an effective theory for the perturbations around an assumed $\Lambda$CDM-like background.  Such an approach has already been taken for models of quintessence \cite{Creminelli:2008wc}, where it was shown that instability issues -- which are common place in such models -- can be addressed within the context of the EFT.  Work in preparation \cite{EFTinprep},  will extend this approach in much the same spirit as here, but with an emphasis on constraining the EFT of the perturbations. This will extend our analysis of cosmological constraints to the EFT of the perturbations, and their implications for complementary constraints from large scale structure measurements of weak lensing, galaxy position and peculiar velocity fields.  These correlations, comparing and contrasting relativistic and non-relativistic tracers, could be a powerful probe of the broad range of gravitational modifications described by the EFT \cite{Zhang:2007nk}. 

%%%%%%%%%%%%%%%%%%%%%%%%%%%%%%%%%%%%%%%%%%%%%%
\section*{Acknowledgments}
%%%%%%%%%%%%%%%%%%%%%%%%%%%%%%%%%%%%%%%%%%%%%%
We would like to thank Jolyon Bloomfield, Eanna Flanagan and Charles Shapiro for useful conversations.
RB's and EM's research is supported by NSF CAREER grant AST0844825, NSF grant PHY0968820, NASA Astrophysics Theory Program grants NNX08AH27G and NNX11AI95G and by Research Corporation. SW would like to thank the George and Cynthia Mitchell Institute for Fundamental Physics, Texas A\&M University, and the Perimeter Institute for hospitality.
%%%%%%%%%%%%%%%%%%%%%%%%%%%%%%%%%%%%%%%%%%%%%

%%%%%%%%%%%%%%%%%%%%%%%%%%%%%%%%%%%%%%%%%%%%%%
% BIB
%%%%%%%%%%%%%%%%%%%%%%%%%%%%%%%%%%%%%%%%%%%%%%

\bibliographystyle{apsrev}

\bibliography{paper}

\end{document}